\pdfoutput=1
\documentclass[12pt]{article}
\usepackage{graphicx,amssymb,amsmath}
\usepackage{epsfig}
\usepackage{color,graphicx}
\usepackage{cite}
\usepackage{amssymb,amsmath}
\usepackage{multirow}
\usepackage{hyperref}
\usepackage[utf8]{inputenc}
\usepackage{subcaption}
\usepackage{tabularx}
\usepackage{float}
\newcolumntype{Y}{>{\centering\arraybackslash}X}
\usepackage[top=2.5cm, bottom=2.5cm, left=2.5cm, right=2.5cm]{geometry}	
\usepackage{slashbox}

\numberwithin{equation}{section}

\newcommand{\be}{\begin{equation}}
\newcommand{\ee}{\end{equation}}
\newcommand{\bea}{\begin{eqnarray}}
\newcommand{\eea}{\end{eqnarray}}

\newcommand{\eps}{\epsilon}

\newcommand{\dd}{\mathrm{d}}

\newcommand{\mt}[1]{\textrm{\scriptsize #1}}
\def\Nc{N_\mt{c}}
\def\Nf{N_\mt{f}}
\def\muB{\mu_\mt{B}}
\def\nB{n_\mt{B}}
\def\nh{\hat{n}}
\def\rH{r_\mt{hor}}  
\def\rHbar{\bar{r}_\mt{hor}}
\def\Vf{V_\mt{f}}
\def\Vg{V_\mt{g}}
\def\sNN{s_\mt{NN}}
\def\ie{{\emph{i.e.}}}
\def\eg{{\emph{e.g.}}}

\def\chione{\chi_1^\mt{B}}
\def\chitwo{\chi_2^\mt{B}}
\def\chithree{\chi_3^\mt{B}}
\def\chifour{\chi_4^\mt{B}}
\def\chisix{\chi_6^\mt{B}}
\def\chieight{\chi_8^\mt{B}}
\def\chiten{\chi_{10}^\mt{B}}
\def\chitwelve{\chi_{12}^\mt{B}}
\def\chin{\chi_n^\mt{B}}
\def\chitwon{\chi_{2n}^\mt{B}}

\begin{document}
 
\begin{flushright}
HIP-2024-13/TH\\
APCTP Pre2024 - 010
\end{flushright}

\begin{center}

\centering{\Large {\bf Refining holographic models of the quark-gluon plasma}}

\vspace{5mm}

\renewcommand\thefootnote{\mbox{$\fnsymbol{footnote}$}}
Niko Jokela,${}^{1,2}$\footnote{niko.jokela@helsinki.fi}
Matti J\"arvinen,${}^{3,4}$\footnote{matti.jarvinen@apctp.org} and
Aleksi Piispa${}^{1,2}$\footnote{aleksi.piispa@helsinki.fi}

\vspace{2mm}
${}^1${\small \sl Department of Physics} and ${}^2${\small \sl Helsinki Institute of Physics} \\
{\small \sl P.O.Box 64} \\
{\small \sl FIN-00014 University of Helsinki, Finland}

 \vskip 0.1cm
 ${}^3${\small \sl Asia Pacific Center for Theoretical Physics} \\
 {\small \sl Pohang 37673, Republic of Korea}

 \vskip 0.1cm
 ${}^4${\small \sl Department of Physics} \\
 {\small \sl Pohang University of Science and Technology} \\
 {\small \sl Pohang 37673, Republic of Korea}

\end{center}

\setcounter{footnote}{0}
\renewcommand\thefootnote{\mbox{\arabic{footnote}}}

\begin{abstract}
\noindent
We investigate quark-gluon plasma at nonzero density by using
two holographic models for QCD based on either brane or Einstein--Maxwell--dilaton actions. We determine the parameters of these models through a systematic statistical fitting procedure to lattice QCD data, which encompasses the equation of state (through the entropy density) and the two lowest baryon number susceptibilities. The predictions from the two models for the higher-order susceptibilities and the equation of state at nonzero density are strikingly similar. In particular, both models suggest the presence of a critical point on the $(\muB,T)$-phase diagram near  $\muB/T \approx 6$. The results for the equation of state are in agreement with lattice data for higher-order susceptibilities and experimental data for the cumulants of the net-proton number from the Beam Energy Scan program at the Relativistic Heavy-Ion Collider. 
\end{abstract}

\newpage
\tableofcontents

\section{Introduction}\label{sec:introduction}

The main physics goal of the Beam Energy Scan (BES) 
program of RHIC is to explore the phase diagram of Quantum Chromodynamics (QCD) at nonzero baryon density, and in particular to 
provide information about the existence of the postulated critical point \cite{Bzdak:2019pkr,Stephanov:1998dy}. Lots of data has already been accumulated, which has probed the properties of the matter in the regime of the crossover from hadrons to quark-gluon plasma \cite{Du:2024wjm}. These results from the first phase of the beam energy scan are limited to relatively small baryon chemical potential, $\muB/T \lesssim 3$~\cite{STAR:2017sal}, but will be extended by the second phase as well as future heavy-ion experiments at FAIR, NICA, and J-PARC to higher densities \cite{Aoki:2021cqa,Lovato:2022vgq,Senger:2022bjo}. Moreover, observations of isolated neutron stars and multi-messenger observations of binary neutron-star mergers provide complementary information about the properties of matter at densities above the nuclear saturation density.

The theoretical description of QCD matter is particularly difficult in the regime of intermediate densities, \ie, densities which are too high for lattice simulations to converge and too low for perturbation theory to work. In the absence of  first principles computational tools one needs to resort to model building in this regime or attempt to extrapolate from available data at low density. The latter method uses lattice QCD data at small baryon chemical potentials $\muB$ and temperatures $T$ around the crossover transition temperature  
and, through educated guess work, maps out the phase diagram in the $(\muB,T)$-plane. A main physics question of such extrapolations is the location of the critical end point (CEP) of the first order deconfinement phase transition on the phase diagram. Note however that even the existence of the critical point is at the moment still unclear. 
On the one hand, current experimental data disfavors QCD critical point in the regime $\muB/T\lesssim 3$. This is also the regime where lattice results can be reliably extended across all available techniques \cite{Philipsen:2005mj,deForcrand:2009zkb,Guenther:2017hnx,Bazavov:2017dus,Ding:2015ona}, showing no signs of a critical point, and moreover there is no strong theoretical reason to expect a deconfinement phase transition to occur even at high densities; see arguments for hadron-quark continuity in \cite{Schafer:1999pb}. On the other hand, the presence of a critical point is predicted by computations in models such as the Nambu--Jona-Lasinio models \cite{Fukushima:2003fw,Ratti:2005jh}.

One of the methods is to model the properties of QCD  
by using a framework that is called the gauge/gravity duality, which is inspired by the more robust string/gravity correspondence \cite{Kim:2012ey,Ramallo:2013bua}. This framework is morally a rewriting of the postulated large-$\Nc$ QCD-like action in terms of classical gravity degrees of freedom and their dynamics. 
Via the gauge/gravity duality, nontrivial questions of strongly coupled field theory can then be mapped to straightforward weakly-coupled computations on the classical gravity side. 
The gravity models typically contain free parameters that are fitted to available data, both from experiments and from lattice simulations. In the context of finite temperature QCD, applications vary from hot QCD heavy-ion phenomena \cite{CasalderreySolana:2011us,Brambilla:2014jmp,Rougemont:2023gfz} to cold and dense neutron-star matter regime \cite{Jarvinen:2021jbd,Hoyos:2021uff}.

We build upon existing  framework to construct a holographic QCD model that fits lattice QCD data at small baryon chemical potential~\cite{Kajantie:2006hv,Gubser:2008ny,Gursoy:2009jd,Misra:2019thm}  and that can then be extrapolated to higher densities~\cite{DeWolfe:2010he,DeWolfe:2011ts}. The original works in this approach appeared long ago
and have been followed and refined by several research collaborations; see reviews \cite{Jarvinen:2021jbd,Hoyos:2021uff,Rougemont:2023gfz} and references therein. In this work we will employ two distinct classes of phenomenological holographic models: Einstein--dilaton gravity coupled to the Dirac--Born--Infeld (DBI) brane action, and Einstein--Maxwell--dilaton (EMD) models. 
The former model is perhaps most faithful to meet expectations from string/gravity correspondence and is exemplified by the V-QCD model~\cite{Jarvinen:2011qe,Alho:2012mh,Alho:2013hsa,Jokela:2018ers} in the context of QCD. Rather than using V-QCD we however use a different approach where the results from the DBI model are compared directly to lattice data at all temperatures, as we will explain in detail below. The EMD models have also been shown to agree, with well chosen potentials, with equation of state (EoS) and baryon number susceptibility $\chitwo$ results from lattice simulations \cite{Critelli:2017oub,Grefa:2021qvt,Li:2023mpv}. 

Apart from working with the DBI action, our work involves novel elements. We develop the fitting procedure towards more systematic method, where we both introduce Ans\"atze for the fitted potentials that can be systematically improved to obtain more accurate fits, and fit the lattice data by using a statistical model. 
As an additional novel component, we simultaneously fit the next baryon number susceptibility, $\chifour(T)$, along with the entropy density and $\chitwo(T)$, to the data provided in \cite{Bazavov:2020bjn}.
In order to do this efficiently, we write a code which computes $\chifour$ in the holographic models from the background data at $\muB=0$ directly, through solving the equations governing the lowest-order fluctuations, therefore avoiding the use of numerical derivatives.

In the subsequent Section~\ref{sec:models}, we will start with a quick overview of the holographic framework by introducing the two types of actions and explain how they can be used to compute equilibrium thermodynamics for given input potentials. These potentials play a key role, as they parametrize our ignorance of the precise gravity dual of QCD. In Section~\ref{sec:modelfitting}, we choose potentials that are motivated by the string/gravity framework and therefore essentially encompass those that are widely used in the literature, but can also be systematically extended to include higher-order corrections. We continue to discuss how these potentials can be fitted to data for the Yang-Mills theory~\cite{Panero:2009tv} and QCD with $\Nf=2+1$ flavors. We use data for entropy density (rather than pressure) because computing entropy density is more efficient in the holographic model. As for the fit to the full QCD data with $\Nf=2+1$ flavors,  we not only use the entropy density of lattice QCD \cite{HotQCD:2014kol,Borsanyi:2013bia} but also carry out a robust fit, determined through least-squares analysis, to the available baryon number susceptibility $\chitwo(T)$~\cite{Bazavov:2020bjn,Borsanyi:2011sw} and $\chifour(T)$~\cite{Bazavov:2020bjn} data. It is important to note that in this paper, quarks are treated as massless, a limitation that we revisit in the conclusion. 

In Section~\ref{sec:results} we present the results from fitting the models to available data. First of all we demonstrate that the results also give an unbiased description of the pressure $P$ by analyzing the interaction measure $(\eps-3P)/T^4$, where $\eps$ is the energy density. We then use the models to compute further susceptibilities and show that they agree well with lattice results which are available up to $\chieight$~\cite{Bazavov:2020bjn}. We provide results for higher-order susceptibilities too. We then continue to nonvanishing density and compare the predictions with the experimental results from the BES program, finding a reasonable matching. We conclude the discussion by mapping out the holographic phase diagram in the ($\muB,T$)-plane. We find a critical point at $\muB/T \approx 6.0$ ($\muB/T \approx 6.4$) in the DBI (EMD) model. 

An interesting outcome of our work is that, despite being distinctly different, the DBI and EMD models produce surprisingly similar quantitative results when extended to large densities. This suggests that holographic modeling of bulk thermodynamic quantities is relatively robust, providing useful qualitative insights. However, caution is needed, as both the DBI and EMD models, in their current formulations, are somewhat simplistic. They do not include chiral symmetry breaking nor confinement, \ie, there are no bound states nor a confinement transition in these models.
In addition, we believe that fitting to flavor-independent lattice data (\ie, only the susceptibilities $\chin$ of the baryon number chemical potential) is insufficient for a precise description of the phase diagram even in the deconfining phase.

We conclude the paper with a detailed discussion in Section~\ref{sec:discussion}, highlighting the limitations of the current holographic modeling. The article is complemented by several appendices containing essential details for reproducing all the results. Additionally, we emphasize the importance of a comprehensive discussion on the underlying invariance of equations of motion under symmetry transformations. Misunderstandings in this area have led to confusion and warrant clarification.

\section{Setup for the holographic models}\label{sec:models}

In this section we will discuss the dual gravity background. We go through salient features while many of the technical details will be relegated in appendices.

\subsection{Actions}\label{sec:action}

We start with laying out the action in the bulk gravity. The aim of this work is bottom-up, hence we will work directly in five dimensions. 
The holographic dictionary includes three fields, the metric $g_{\mu\nu}$, the dilaton $\phi$, and the gauge field $A_\mu$.
Per usual gauge/gravity dictionary, the gauge potential is dual to a $U(1)$ current in the field theory, which here is identified as the baryon number current. The source for the temporal component of the gauge field is the baryon chemical potential $\muB$.
The boundary value of the scalar field $\phi$ sources the 't Hooft coupling, dual of the operator $\mathrm{Tr}{\cal{F}}^2$ with ${\cal F}$ the field strength of the gauge theory, while the five-dimensional metric $g_{\mu\nu}$ is dual to the energy-momentum tensor of field theory and sources the metric on the field theory side (which we take to be the Minkowski metric with mostly plus convention). 

We consider models in two classes, Einstein-dilaton gravity with DBI matter sector, and EMD models.
The action for the DBI model is
\be\label{eq:actionDBI}
 S_{\mathrm{DBI}} = \frac{1}{2\kappa_5^2} \int \dd^5 x \bigg\{ \sqrt{-g}\bigg[R-\frac{1}{2}g^{\mu\nu}\partial_\mu\phi\partial_\nu \phi - \Vg(\phi) \bigg] - \Vf(\phi)\sqrt{-\mathrm{det}(g_{\mu\nu}+w(\phi)F_{\mu\nu})}\bigg\} \ .
\ee
Here the Newton constant is given by $G_5=\kappa_5^2/(8\pi)$,  $R$ is the scalar curvature of the  metric $g_{\mu\nu}$, 
and  $F_{\mu\nu}$ is the field strength of the Abelian bulk gauge field $A_\mu$. The potentials $\Vg$, $\Vf$, and $w$ will be determined below by comparing to QCD data.
The action~\eqref{eq:actionDBI} consists of two parts, the first term (Einstein-dilaton gravity) describes the gluonic sector while the second term is associated with flavor degrees of freedom. This particular action is in the DBI form, closely following the stringy description for flavor branes \cite{Karch:2002sh}.

The action for the EMD model is given by
\be\label{eq:actionEMD}
 S_{\mathrm{EMD}} = \frac{1}{2\kappa_5^2} \int \dd^5 x\sqrt{-g}\bigg[R-\frac{1}{2}g^{\mu\nu}\partial_\mu\phi\partial_\nu \phi - V(\phi) -\frac{Z(\phi)}{4}F_{\mu\nu}F^{\mu\nu}\bigg] \ .
\ee
That is, here  the flavor sector is described by a Maxwell term rather than the DBI action. The potentials $V$ and $Z$ will be determined by fitting to lattice data below. The action (\ref{eq:actionEMD}) can be obtained from (\ref{eq:actionDBI}) in the limit of slowly varying fields or in the present context at low densities, which operatively means developing the DBI action as a series in the field strength tensor $F_{\mu\nu}$.
That is, at small baryon chemical potential the two actions coincide if we identify
\be\label{eq:DBIEMDidentification}
 V(\phi)=\Vg(\phi) + \Vf(\phi) \ , \qquad Z(\phi) =  \Vf(\phi) w(\phi)^2 \ , \qquad (\muB\to 0) \ .
\ee

In the above actions, boundary terms including the Gibbons--Hawking--York term and regularization terms for divergence cancellation for on-shell quantities such as obtaining the equation of state are omitted for brevity, but will be briefly returned to later in the text.

\subsection{Background}\label{sec:background}

We are interested in homogeneous and isotropic field configurations. Therefore our Ansatz for the metric only depends on the bulk holographic coordinate $r$:
\be
  \dd s^2  =  e^{2A(r)}\left(-h(r)\dd t^2+\dd\mathbf{x}^2\right) + \frac{e^{2B(r)}}{h(r)}\dd r^2 \ .
\ee
Here $h(r)$ is the blackening factor which vanishes at the horizon of the black hole, $h(\rH)=0$. Such a horizon is always present in our geometries.  One of the functions $A(r)$ and $B(r)$ can be determined through leftover parameterization freedom. Typical choices include $A(r)=B(r)$ (conformal coordinates) and $B(r)=0$ (domain wall coordinates). We will use here the latter choice.   In these coordinates the boundary is located at coordinate infinity, and we choose the direction of the flow such that the boundary is at positive values of the coordinate, \ie, at $r \rightarrow \infty$. We also consider non-trivial scalar field $\phi$ and turn on a temporal component of the gauge field:
\be\label{eq:ansatz}
  \phi  =  \phi(r)  \ , \qquad A_\mu \dd x^\mu = \psi(r)\dd t \ , 
\ee
where the latter turns on the baryon number chemical potential through the holographic dictionary,
\be \label{eq:muBdef}
 \lim_{r\to \infty}\psi(r) = \muB \ .
\ee
According to the dictionary, the source of the dilaton field is the coupling of QCD. However both in the field theory and the holographic models the coupling runs, and the coupling is converted into an energy scale, which we denote by $\Lambda$ in the holographic models (and $\Lambda_\mathrm{QCD}$ in QCD). This scale can be defined through the near boundary asymptotics of the dilaton after we have specified the potentials in the action, see discussion below in Sec.~\ref{sec:potentials}.
Since we do not introduce a field dual to the chiral condensate, chiral symmetry is intact, which implicitly also means that we are working in the limit of massless quarks.

We relegate the full equations of motion following from actions (\ref{eq:actionDBI}) and (\ref{eq:actionEMD}) in Appendix~\ref{app:backgroundEOM} for completeness. Before continuing, however, let us note that, in the absence of  
any Wess--Zumino terms in the action, the gauge potential $\psi$ is a cyclic variable with the Ansatz~\eqref{eq:ansatz}. This means that we can simply obtain its first integral:
\be\label{eq:conservedcharge}
    \frac{2\kappa_5^2}{V_4}  \frac{\delta S_\mathrm{DBI}}{\delta \psi'} = \nh = \frac{\Vf(\phi)w(\phi)^2e^{2A-B}\psi'}{\sqrt{1-w(\phi)^2e^{-2(A+B)}\psi'^2}} \ ,
\ee
where $V_4$ is the volume of spacetime, $\nh$ is a constant of integration, and the prime denotes differentiation with respect to $r$. We wrote here the expression for the DBI action, for the EMD case one simply drops the square root factor and applies the mappings  
(\ref{eq:DBIEMDidentification}) (see Appendix~\ref{app:emd}). From~\eqref{eq:conservedcharge} we directly note that the integration constant $\nh$ is related with the conserved quantity, \ie, that the expression on the right hand side does not depend on $r$.

\subsection{Thermodynamics}\label{sec:cumulants}

Our bulk geometry always has a black hole. This means that we are only  describing the field theory in the deconfining phase, which extends to all nonzero temperatures and densities and covers the whole phase diagram, 
following the approach of~\cite{Gubser:2008ny,DeWolfe:2010he}. The temperature $T$ of which is identified with the temperature of the black hole given by the surface gravity, extracted from
\be \label{eq:physT}
 T = \frac{e^{A(\rH)}}{4\pi}h'(\rH) \ .
\ee
In addition, the entropy density of the field theory is identified with the Bekenstein--Hawking entropy of the black hole, obtained from computing its area,
\be\label{eq:physs}
 s = \frac{2\pi}{\kappa_5^2}e^{3A(\rH)} \ .
\ee
The baryon number chemical potential $\muB$ is given as the boundary value of the gauge field, see~\eqref{eq:muBdef}. The constant $\nh$ is proportional to the baryon charge density in the field theory:
\be 
 \nB = \frac{\nh}{2\kappa_5^2} \ .
\ee

The dimensionless susceptibilities $\chitwon(T)$ at zero baryon chemical potential $\muB=0$ are defined as the derivatives of the pressure with respect to the chemical potential
\be \label{eq:chi2ndef}
 \chitwon(T) = T^{2n}\frac{\partial^{2n}(P(T,\muB)/T^4)}{\partial\muB^{2n}}\bigg{\lvert}_{T\ \mathrm{fixed},\ \muB = 0} \ .
\ee
Due to charge conjugation symmetry, only even powers of $\muB$ contribute to the pressure. A natural extension for calculating the susceptibilities at nonzero baryon chemical potential is given by
\be \label{eq:chindef}
 \chin(T,\muB) = T^{n}\frac{\partial^{n}(P(T,\muB)/T^4)}{\partial\muB^{n}}\bigg{\lvert}_{T\ \mathrm{fixed}} \ .
\ee
At nonzero chemical potential, odd-order susceptibilities are generally nonzero as well.

The definition~\eqref{eq:chi2ndef} implies the following series expansion for the pressure 
\be \label{eq:pressure}
    \frac{P(T,\muB)-P(T,\muB=0)}{T^4} = \sum_{n=1}^{n_\text{max}}\frac{1}{(2n)!}\chitwon(T)\bigg(\frac{\muB}{T}\bigg)^{2n} +\mathcal{O}\left[\bigg(\frac{\muB}{T}\bigg)^{2(n_\text{max}+1)}\right]\ ,
\ee
\ie, we are assuming analyticity. There is an associated convergence radius to this series expression presumably set by the critical point \cite{Wen:2024hbz}. In principle, one should be concerned about the Lee--Yang singularities that will follow. We will effectively work with the truncated series where this is not an issue ($n_\text{max}<\infty$); see~\cite{Basar:2023nkp} for recent work on resummations. 
The baryon charge density is defined as the first derivative of the pressure with respect to the baryon chemical potential 
\be \label{eq:rhoexp}
    \frac{\nB(T,\muB)}{T^3} =T \frac{\partial(P(T,\muB)/T^4)}{\partial\muB}\bigg|_{T} =  \sum_{n=1}^{n_\text{max}} \frac{1}{(2n-1)!} \chitwon(T)\bigg(\frac{\muB}{T}\bigg)^{2n-1} +\mathcal{O}\left[\bigg(\frac{\muB}{T}\bigg)^{2n_\text{max}+1}\right]\ .
\ee

In the holographic DBI model~\eqref{eq:actionDBI} we obtain a relation between the chemical potential and the normalized density $\nh =2\kappa_5^2 \nB$:
\be \label{eq:Idef}
 \frac{\muB}{\nh} =  \int_{\rH}^\infty \dd r\ \frac{1}{e^{2A}\Vf(\phi)w(\phi)^2}\left(1+\frac{\nh^2}{e^{6A}\Vf(\phi)^2w(\phi)^2}\right)^{-\frac 12} \ .
\ee
See Appendix~\ref{app:backgroundEOM} for details.
Expanding this integral as a series (at fixed temperature),
\be \label{eq:I2ndefs}
 \frac{\muB}{\nh} = \sum_{n=0}^{\infty} I_{2n} \nh^{2n}
\ee
formally defines the factors $I_{2n}$. Notice that this definition is not a direct Taylor expansion of the explicit $\nh$ dependence in~\eqref{eq:Idef}, because the functions $A$, $\phi$, and the location of the horizon $\rH$ depend implicitly on $\nh$.

Inserting the expansion~\eqref{eq:I2ndefs} in~\eqref{eq:rhoexp} we find the following formulas for the first two dimensionless susceptibilities 
\bea
    \chitwo(T) & = & \frac{1}{2\kappa_5^2}\frac{1}{T^2I_0}  \label{eq:chitwo}\\
    \chifour(T) & = & -\frac{1}{2\kappa_5^2}\frac{6I_2}{(I_0)^4} \label{eq:chifour}\ . 
\eea
The leading coefficient $I_0$ can be immediately written as an explicit integral
\be
 I_0 = \int_{\rH}^\infty \dd r\ \frac{1}{e^{2A}\Vf(\phi)w(\phi)^2}
\ee
which is to be evaluated at zero density. For $\chitwo(T)$ the result can therefore  be written in terms of a single integral over the background solution
\be \label{eq:chi2int}
    \chitwo(T) = \frac{1}{2\kappa_5^2T^2}\frac{1}{\int_{\rH}^\infty \dd r \ e^{-2A}\Vf(\phi)^{-1} w(\phi)^{-2}} \ .
\ee
For the Einstein--Maxwell-dilaton model the construction works with slight changes: in~\eqref{eq:Idef}  
the square root factor should be dropped (see Appendix~\ref{app:emd}). Nevertheless, the expression for $\chitwo$ takes the same form, given in~\eqref{eq:chi2int}, upon identifications in (\ref{eq:DBIEMDidentification}). 

Similar formulas as~\eqref{eq:chitwo} and~\eqref{eq:chifour} can also be derived for the higher-order susceptibilities (see Appendix~\ref{app:cumulants}), leading to the following observation: The equation (\ref{eq:Idef}), 
along with the analogous formula for the EMD action, contains a wealth of information. By solving this equation for baryon charge density, one can directly express it as a function of chemical potential {\emph{and}} temperature. Therefore,  
one could extract all order susceptibilities from (\ref{eq:Idef}) at any point in the phase diagram and, in particular, contemplate fitting the holographic model to all available susceptibilities from lattice computations at vanishing density. Apart from the lowest order susceptibility, this is however not straightforward.
This is because only the lowest order coefficient $I_0$, and consequently only the lowest order susceptibility $\chitwo(T,\muB=0)$, can be obtained in closed form by using this relation.
The knowledge about higher-order susceptibilities requires taking into account the implicit dependence of the background solution for the geometry and the dilaton on the charge in the relation~\eqref{eq:Idef}. 
The simplest approach is to solve the integral around a specific chemical potential at fixed temperature and then compute numerical derivatives. However, this method is numerically costly and tends to produce increasing errors as the order of susceptibility rises.
We will show how to overcome the hindrance of computing numerical derivatives by instead fluctuating the background for the next integral $I_2$ and the susceptibility $\chifour$:  
in order to derive an analytic formula for the factor $I_2$ at vanishing density it is enough to consider the leading perturbation of the background around $\nh=0$. We carry out this analysis explicitly in Appendix~\ref{app:fluctuations}.

With the holographic dictionary we can calculate many thermodynamic variables using standard thermodynamic relations. Through the Smarr formula \cite{Karch:2015rpa} in (\ref{eq:charge2}), the internal energy density $\epsilon$ and the pressure $P$ satisfy the expected relation
\be
    \epsilon+P  =  Ts+\muB\nB \ .
\ee
Varying the on-shell action one obtains the fundamental relation 
\be
    \dd P = s\dd T+\nB \dd\muB
\ee
which upon integration will give the pressure:
\be\label{eq:Pintegration}
    P(T,\muB=0) = \int_{T_\text{min}}^{T}\dd\tilde{T} s(\tilde{T},\muB=0) \ .
\ee
Note, however, that there is an integration constant, \ie, one needs to specify the zero of the pressure for instance. This is linked with the lower integration limit $T_\text{min}$. In our setup the  standard choice would be $T_\text{min} = 0$ but as it turns out, solving the required black hole backgrounds at low temperatures is numerically challenging.  
We will choose $P(T_\text{min},\muB=0)=0$ for the lowest $T_\text{min}$ that our numerical methods allow (this is quantitatively vindicated below). 
Having obtained the pressure from above integration, we can then get an expression for the integration measure at zero density, \be\label{eq:interactionmeasure}
 \epsilon-3P = Ts 
 -4P \  \ \ \ \ , \ \ (\muB=0) \ .
\ee

\section{Fitting the models to lattice data}\label{sec:modelfitting}

In this section we will describe the fitting procedure. First we discuss how the potentials are parametrized. We then explain how the fitting procedure can be done in stages, where the parameters in different potentials are independently fitted to different sets of observables.

\subsection{Parametrization of potentials}\label{sec:potentials}

In this subsection we will choose a specific parametrization of our potentials appearing in the actions (\ref{eq:actionDBI}) and (\ref{eq:actionEMD}).

Let us first discuss how we choose the behavior of the geometry in the weak coupling regime, and what this implies for the potentials. That is, QCD becomes asymptotically free at high energies, \ie, the theory flows to a trivial conformal UV fixed point where the coupling tends to zero and interactions are suppressed.  Since holography with classical gravity makes only sense at strong coupling, there are no strict guidelines for choosing  weak coupling behavior. A possible choice is to adjust the action such that $\phi \to -\infty$ at the boundary and identify $e^{\phi}$ as the 't Hooft coupling in QCD, which therefore tends to zero. This is done in the improved holographic QCD (IHQCD) framework~\cite{Gursoy:2007cb,Gursoy:2007er}, where, in addition, the potential of the dilaton gravity is chosen such that $e^{\phi}$ runs logarithmically to zero, mimicking the renormalization group flow of QCD. Another popular choice is to choose a dilaton potential with a maximum at finite value of the coupling, in analogy to the asymptotic safety picture. In both cases, the geometry is asymptotically AdS$_5$ at the boundary (see Appendix~\ref{app:backgroundEOM}). We will follow here the latter choice. 

Therefore our dilaton potential $V$ (or the combination $\Vg+\Vf$ for the DBI case with flavors) satisfies
\be \label{eq:UVpot}
 V(\phi) = -\frac{12}{L^2} + m^2 (\phi-\phi_0)^2 + \mathcal{O}\left[ (\phi-\phi_0)^3\right] \ ,
\ee
where $L$ is the asymptotic AdS$_5$ radius. We will set the minimum value $\phi_0$ to zero. The mass of the dilaton $m$ is related to the weak-coupling dimension of the operator $\mathrm{Tr} {\cal{F}}^2$ through
\be \label{eq:Deltadef}
 \Delta (4-\Delta ) = - m^2 L^2 \ .
\ee
The perturbative value of the dimension is $\Delta = 4$, but the renormalization group flow drives it to slightly lower values as one flows away from the fixed point. 
Thus,  
realistic choices of $\Delta$ will be close but slightly below 4. We will treat it below as an independent fitting parameter, while other works have found best to keep it fixed, see, \eg, \cite{Li:2023mpv,Liu:2023pbt}.

The choice of the UV behavior~\eqref{eq:UVpot} also determines the asymptotics of the dilaton and the geometry near the boundary, see Appendix~\ref{app:backgroundEOM}. This allows us to give a precise definition for the scale $\Lambda$ of the renormalization group flow of the background, identified through the source term of the dilaton: 
\be \label{eq:Lambdadef}
 \Lambda = \lim_{r\to\infty} e^{A(r)}\phi(r)^{1/(4-\Delta)} \ .
\ee

The strong coupling behavior of the potentials will be determined by the lattice data. For the Ans\"atze we however choose functions that behave exponentially as is typical for potentials in string theory. It has also been demonstrated that the limiting critical behavior between confining and deconfined vacua in Einstein-dilaton gravity is achieved using an exponential potential with a specific coefficient in the exponent~\cite{Gursoy:2007cb}. Potentials chosen in the earlier literature~\cite{Gubser:2008yx,Gubser:2008ny,Gursoy:2008bu,Gursoy:2008za} are typically exponentials with coefficients close to this critical value. Apart from exponential terms, in each function we add a polynomial, so that the functions can be systematically extended to give an arbitrarily precise description of given data by increasing the order of the polynomial. Note that our fit functions are generic, as no specific features are imposed, aside from the exponential asymptotics.  
In our approach the fit functions typically have a rather large number of parameters. It is also possible to adjust the fitted functions such that the number of parameters is dramatically reduced (see Appendix~\ref{app:reduction}) but this comes at the expense of losing the systematic expandability.
There is also a simple hierarchy concerning the coefficients  of the polynomial in our functions and the temperature ranges in our fits. Because lower temperatures in our setup are obtained for smaller black holes, the values of the dilaton near the black hole horizon grow with decreasing $T$~\cite{Gubser:2008ny}. Therefore terms with higher (lower) powers in our polynomial Ansatz mostly control the thermodynamics at low (high) temperatures.

Let us then discuss explicitly the choices for the DBI action (\ref{eq:actionDBI}). The DBI term models the quark degrees of freedom. Hence, to obtain the pure glueball dynamics we can set $\Vf(\phi) = 0$. We will choose the potential governing the glueball dynamics in close analogy to \cite{DeWolfe:2010he,DeWolfe:2011ts} as follows 
\be \label{eq:Vgpot}
 \Vg(\phi)  =  -\frac{12}{L_0^2} \cosh(\gamma_1 \phi) + v_2 \phi^2+v_4 \phi^4+v_6 \phi^6 \ ,
\ee 
where $L_0$ is the radius of curvature in the absence of flavors, \ie, for $\Vf(\phi)=0$, and where $\gamma_1$ and $v_i$ are fit parameters. 
This Ansatz can be systemically extended by adding higher-order polynomial terms, including odd powers $\phi^{2k+1}$ with $k\ge 1$. Notice that the linear term is excluded due to the choice of weak-coupling behavior in~\eqref{eq:UVpot}.
This potential is fitted to pure large-$\Nc$ $SU(\Nc)$ Yang--Mills data. The details of this fitting is discussed below in Sec.~\ref{sec:dataselection}. This fitting fixes the parameters shown in the potential and the Newton constant $\propto\kappa_5^2$.

In the limit of vanishing baryon chemical potential all components of the gauge fields vanish, so that  $F_{\mu\nu}=0$. Thus at zero chemical potential the dilaton potential of the full DBI model is given by $\Vg(\phi)+\Vf(\phi)$. To preserve the UV form of the potential, that is, cosmological constant and a mass term, we choose the following Ansatz for $\Vf(\phi)$
\be\label{eq:Vfpot}
    \Vf(\phi) = f_0 + f_2 \phi^2 + f_4\phi^4 + f_6 \phi^6 \ .
\ee
In particular, the coefficients $f_0$ and $f_2$ modify the values of the cosmological constant and the dilaton UV mass term: 
the coefficient $f_0$ models the difference in the number of degrees of freedom between QCD and pure Yang--Mills and the coefficient $f_2$ models the effect of  the backreaction of quarks to scaling dimension of $\mathrm{Tr}{\cal{F}}^2$. This potential is fixed by fitting the model to $\Nf=2+1$ flavor lattice QCD at zero chemical potential. Moreover, this fit fixes the energy scale $\Lambda$ of~\eqref{eq:Lambdadef} (see Appendix~\ref{app:compdetails}). 

The inclusion of the gauge field in the DBI action adds one more potential that we call $w(\phi)$. This potential governs the physics at nonzero density. From string theory we expect this potential to falloff as an exponential for large values of the dilaton.  This motivates the Ansatz, inspired by \cite{DeWolfe:2010he,DeWolfe:2011ts},
\be\label{eq:wpot}
    w(\phi)^{-1} = w_0\cosh(\gamma_2 \phi) + w_2\phi^2 + w_4\phi^4 + w_6\phi^6 \ .
\ee
This potential is fixed by fitting it to susceptibility data from lattice QCD. As above, we could add odd powers in $\phi$ in the potential (\ref{eq:wpot}). In this case, also a linear term is allowed because $\phi=0$ does not need to be an extremum of $w(\phi)$.

In the EMD case, we make choices that are analogous to our choices for the DBI action: 
\bea
  V(\phi) & = & -\frac{12}{L_0^2} \cosh(\tilde\gamma_1 \phi) + b_2 \phi^2+b_4 \phi^4+b_6 \phi^6  \label{eq:Vpot} \\
    Z(\phi)^{-1} & = & z_0\cosh(\tilde\gamma_2 \phi) + z_2\phi^2 + z_4\phi^4 + z_6\phi^6  \label{eq:Zpot} \ .
\eea
As suggested by the identification~\eqref{eq:DBIEMDidentification}, the potential $V(\phi)$ controls the equation of state of the full flavored QCD, and will be fitted to the lattice data for entropy density, while the function $Z(\phi)$ controls the dependence on chemical potential, and will be fitted to the susceptibilities. 
In order to make the comparisons between the EMD and DBI setups easier, we also require that the UV dimensions defined via~\eqref{eq:Deltadef} are equal. Then the leading order flows in the two setups match near the boundary. Using the expansion~\eqref{eq:UVpot} with the Ans\"atze~\eqref{eq:Vgpot},~\eqref{eq:Vfpot}, and~\eqref{eq:Vpot}, this implies
\be
 b_2-6\tilde{\gamma}^2_1/L_0^2 = \frac{v_2+f_2-6\gamma^2_1/L_0^2}{1-f_0 L_0^2/12}
\ee
which we use to eliminate one of the EMD parameters, say $b_2$. Notice that the value of $\Delta$ in the DBI and EMD fits would be similar even without imposing this condition, and the condition constrains the potentials near the boundary, whereas the thermodynamics arises mostly from the geometry deeper in the holographic direction. Therefore imposing this condition is expected to have a negligible effect on our fits.
We also remark that comparing to the lattice data for pure Yang-Mills theory is irrelevant for the EMD model. Naturally, one could fit the $V(\phi)$ above also to Yang-Mills data, but this would be a separate fit, which would not affect the final model.   

Moreover, note that we also fit the parameter $z_0$ to data. In principle, one could without loss of generality set $z_0=Z(0)^{-1}$ to one, as this normalization can be obtained by rescaling the gauge field in the EMD action~\eqref{eq:actionEMD}. Such rescaling would however affect the units of the baryon number chemical potential through~\eqref{eq:muBdef}. Some earlier fits in the literature indeed set $Z(0)=1$ (approximately or exactly, see~\cite{Knaute:2017opk,Critelli:2017oub,Cai:2022omk,Liu:2023pbt,Hippert:2023bel,Fu:2024wkn}), but apparently do not explicitly take into account the change of units of the chemical potential and charge when fitting to the $\chitwo$ data. This leads to a fit for the function $Z(\phi)$ which has a prominent, narrow spike near $\phi=0$. However, as the coupling functions ($w(\phi)$ for DBI and $Z(\phi)$ for EMD) of the gauge field only appear in the formula for $\chitwo$ through an integral, see~\eqref{eq:Idef} and~\eqref{eq:munBrelEMD} in Appendix~\ref{app:emd}, the effect of the spike on the result is essentially negligible. We discuss this in more detail in Appendix~\ref{app:reduction}. We stress however that the fits in these references are done correctly, and only remark that the spike can be dropped without essentially changing the results, which simplifies the fit.

Lastly, we will briefly discuss how our current approach with the DBI action contrasts with the previous practice of fitting V-QCD to lattice data~\cite{Gursoy:2009jd,Jokela:2018ers}. 
Namely, the action of V-QCD in the deconfined phase, assuming similar setup as here (only baryonic chemical potential and chirally symmetric phase with zero quark masses so that all quark flavors are alike) coincides with our DBI model in~\eqref{eq:actionDBI}. The main difference, that already appears in the IHQCD model~\cite{Gursoy:2007cb,Gursoy:2007er} that describes the pure glue sector of QCD is that there is a deconfinement phase transition. Therefore the vacuum in V-QCD is confining while in the approach~\cite{Gubser:2008ny,DeWolfe:2010he} that we are following here it is not. This is why the V-QCD model is fitted to lattice data for QCD thermodynamics only in the deconfined phase, \ie, for temperatures above the crossover scale~\cite{Jokela:2018ers,Jarvinen:2022gcc}. Moreover, the confining phase in the V-QCD model is chirally broken, which is implemented on the gravity side through a condensation of an additional scalar field~\cite{Bigazzi:2005md,Casero:2007ae,Dhar:2007bz,Bergman:2007pm,Jokela:2009tk}.

The potential selections in the V-QCD scenario are more complex for two reasons. Firstly, the analog glue potential $\Vg(\phi)$ that was found to work well with backreacted flavors \cite{Alho:2015zua} obeys a specific asymptotics at small coupling, which is 
crucial for accurately capturing running coupling effects up to two loops in the UV. Secondly, reflecting the above discussion about confinement, there is a judiciously chosen combination of a power-law times an explicit logarithm that gives a (good-type \cite{Gubser:2000nd}) singularity of the metric in the IR to achieve confinement and asymptotically linear glueball spectrum in the confined phase~\cite{Gursoy:2010fj}. The flavor potentials \cite{Jokela:2018ers}, analog of $w(\phi)$, were chosen not to spoil these qualities, agree with additional features in the quark sector (see the review~\cite{Jarvinen:2021jbd}), before fitting to lattice data in the same way as we do here.

\subsection{Data selection and fits}\label{sec:dataselection}

Let us then discuss how the fitting procedure works and to which data we are fitting.

As alluded to before, we do not wish to fit the pressure, or equation of state, although accurate data does exist for it. Instead of pressure we fit lattice data for its derivative with respect to temperature, \ie, the entropy density. 
The main reason is that computing the pressure in the holographic models is tricky. It is given as the value of the on-shell action, which however needs to be regularized, and computing its value directly turns out to be numerically challenging albeit possible. It is actually easier to obtain results for the pressure by first computing entropy and integrating with respect to temperature. However this means that whenever the parameters are changed, the entropy needs to be computed in a range of temperatures to obtain pressure at a single value. 

In addition, note that because the pressure is divergent, it is also scheme dependent. The scheme dependence however only affects an overall constant term, and it can be avoided by considering difference of pressures between the vacuum and the state at nonzero temperature and density, as also obtained by considering at integral over the entropy as we did above in Sec.~\ref{sec:cumulants}.
Apart from directly  fitting the entropy, novel components in our methodology include a statistical approach as well as incorporating higher-order baryon number susceptibilities, as will be detailed below.

The fit will be carried out in stages. The first step, which is only present in the DBI model, is to determine the pure glue potential $\Vg$. Since we are eventually interested in the thermodynamics of full QCD with quarks included, the choice of $\Vg$ is not expected to play a major role for the final results. The fits of the other potentials in the subsequent steps will be much more important for the predictions we present later on. Note also that, unlike full QCD with physical quark masses, pure Yang--Mills theory exhibits a first-order deconfinement phase transition at nonzero temperature. However as we are following the approach of~\cite{Gubser:2008ny,DeWolfe:2010he}, we will fit the Yang--Mills data with a model that has no deconfinement transition. This leads to some compromises in the quality of the fit, as we show below.

In the next two stages, common to both the DBI and EMD models, we first fit the model at zero density to the entropy density data, and then fit the functions governing nonzero density behavior (\ie, gauge field couplings) to the susceptibility data.
Note that we are only using data with $\Nf=2+1$ flavors with physical quark masses in these last two steps. This is because holographic models only contain massless quarks, and therefore is unlikely to correctly describe the effect due to heavy quarks. That is, we expect that most of the effects of the light quark masses and the strange quark mass can be accounted for effectively simply by adjusting the potentials, rather than including the quark masses explicitly following the holographic dictionary.  Note that there is also data for two-flavor QCD \cite{Karsch:2001vs,Allton:2005gk} but this is available only for unphysically large quark masses.

\begin{figure}[hbt!]
\centering
\begin{subfigure}{.5\textwidth}
  \centering
  \includegraphics[width=\linewidth]{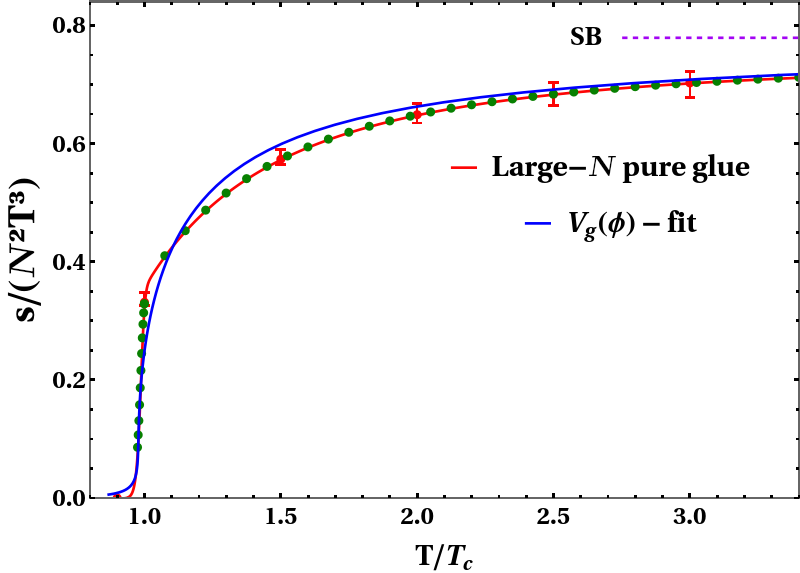}
  \caption{}
  \label{fig:spureglue}
\end{subfigure}%
\begin{subfigure}{.5\textwidth}
  \centering
  \includegraphics[width=\linewidth]{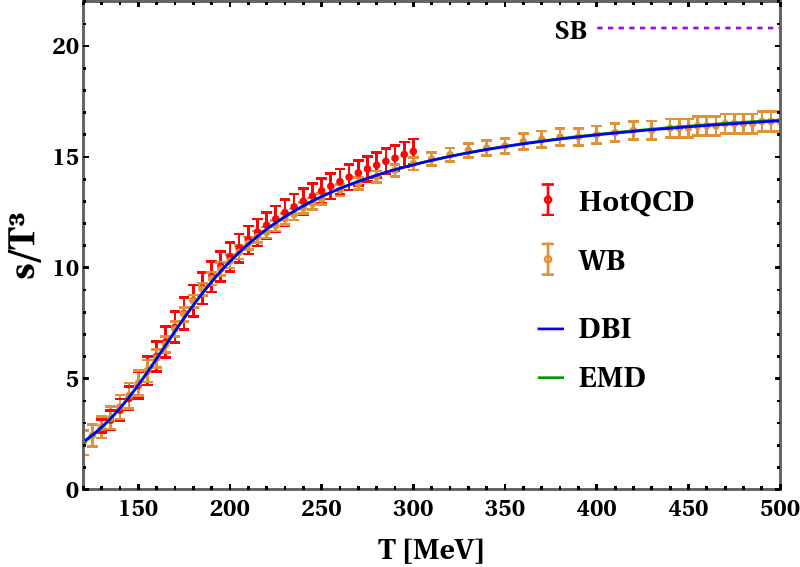}
  \caption{}
  \label{fig:s}
\end{subfigure}
\caption{Fits to the entropy density data.
{\bf Left}: Fit to the large-$\Nc$ pure glue entropy density of \cite{Panero:2009tv}, shown as the red curve and green dots, respectively. Our model fit is shown as the blue curve. 
{\bf{Right}}: $\Nf=2+1$ flavor QCD entropy density of \cite{Borsanyi:2013bia} (orange) and \cite{HotQCD:2014kol} (red). The blue (green) curve corresponds to the holographic fit of the DBI (EMD) model to the continuum entropy density data.  
SB refers to the Stefan--Boltzmann limit discussed in Appendix~\ref{app:inftemp}.}
\label{fig:entropies}
\end{figure}

\paragraph{Pure glue:}
The data for the pure glue fixes the potential $\Vg(\phi)$ in the DBI model. 
For the fit we use the large-$\Nc$ entropy data with $\Nc=3$ from \cite{Panero:2009tv}. Error bars are not taken into account because the
errors were only listed at six different reference points in this reference. The errors show relatively weak dependence on temperature, see the red points in Fig.~\ref{fig:spureglue}.
We use 44 data points, denoted as green dots in Fig.~\ref{fig:spureglue}, from \cite{Panero:2009tv}. The distribution of the points was chosen such that the resulting fit is reasonable near the transition at $T=T_c$.
The measure for the goodness of the fit is therefore chosen to be 
\be\label{eq:gluechi2}
\chi^2_\mt{glue} = \frac{1}{n}\sum_{i=1}^n \frac{(X_i-X^\mathrm{fit}_i)^2}{X_i^2} \ .
\ee
Here $n$ is the size of the used data set, $X_i$ label the data points, and $X^{\mathrm{fit}}_i$ are the corresponding computed points from the fit of the holographic model.
Note that the measure (\ref{eq:gluechi2}) weighs strongly the region where the entropy is small. This region is harder to capture with the Ansatz considered here as there is a phase transition. 

The large-$\Nc$ pure glue entropy density data of \cite{Panero:2009tv} used for the fit is shown in Fig.~\ref{fig:spureglue} as green dots and the red curve as well as the red error bars.
Our fit result is shown as the blue curve. 
As expected, there is a small tension between the data and the fit near $T=T_\mt{c}$, because the holographic model does not contain a phase transition.
The fit parameters, \ie, the parameters of the $\Vg(\phi)$ potential and the Newton constant $\propto \kappa_5^2$ are given in Table~\ref{tab:fitresultsDBI}. 
The high-temperature Stefan--Boltzmann limit discussed in Appendix~\ref{app:inftemp} is shown as the dashed magenta line. Notice that the limit in \cite{Panero:2009tv} is different from ours as we show here the  
result for the entropy density of 
gas of $\Nc^2-1$ gluons with $\Nc=3$ whereas \cite{Panero:2009tv} shows the leading term $\propto \Nc^2$ at large $\Nc$.

\paragraph{Entropy density:}
The data for the QCD entropy density fixes the potential $\Vf(\phi)$ for the DBI model and $V(\phi)$ for the EMD model (see Appendix~\ref{app:emd}). 
\begin{itemize}
 \item HotQCD: We use the entropy density data from \cite{HotQCD:2014kol} for the fit. We will choose the fitted data points as follows: the whole range of available data in steps of 10 MeV, from 130 MeV to 300 MeV, taking into account the error bars. The data is in steps of 5 MeV, but in order to make a fully consistent comparison with the WB data, we will increase the step size to 10 MeV. This data is shown in Fig.~\ref{fig:s} in red. 
 \item Wuppertal--Budapest: \cite{Borsanyi:2013bia} 
 has $\Nf=2+1$ flavor entropy density data which we also take into account.  We will choose the fitted data points as follows: in steps of 10 MeV, from 110 MeV to 400 MeV, taking into account the error bars, and dropping the data in the range from 410 MeV to 510 MeV. This is because as the temperature grows 
we enter the region where the thermodynamics is determined by weakly coupled physics and falls beyond the scope of the strongly coupled holographic description. 
This data is shown in Fig.~\ref{fig:s} in orange. 
\end{itemize}

We introduce a statistical measure on the goodness of the fit as follows:
\be\label{eq:chisquared}
\chi^2_\mt{entropy} = \frac{1}{n} \sum_{i=1}^n  \frac{(X_i-X^\mathrm{fit}_i)^2}{\Delta X_i^2} \ ,
\ee
where $n$ is the size of the used data set. Note that here, in contrast to (\ref{eq:gluechi2}), we are taking into account the errors $\Delta X_i$ from lattice data. 
When the error is not symmetric, we use the larger value.

The fits for the DBI (EMD) model are shown as the  blue (green) curve in Fig.~\ref{fig:entropies}. The green curve is barely visible because it almost overlaps with the blue curve. These fits fix the potential $\Vf(\phi)$ in DBI, and $V(\phi)$ as well as the Newton constant $\propto \kappa_5^2$ in EMD. The fitted parameter values are given in Tables~\ref{tab:fitresultsDBI}~and~\ref{tab:fitresultsEMD} for DBI and EMD, respectively.

\begin{figure}[H]
\centering
\begin{subfigure}{.49\textwidth}
  \centering
  \includegraphics[width=\linewidth]{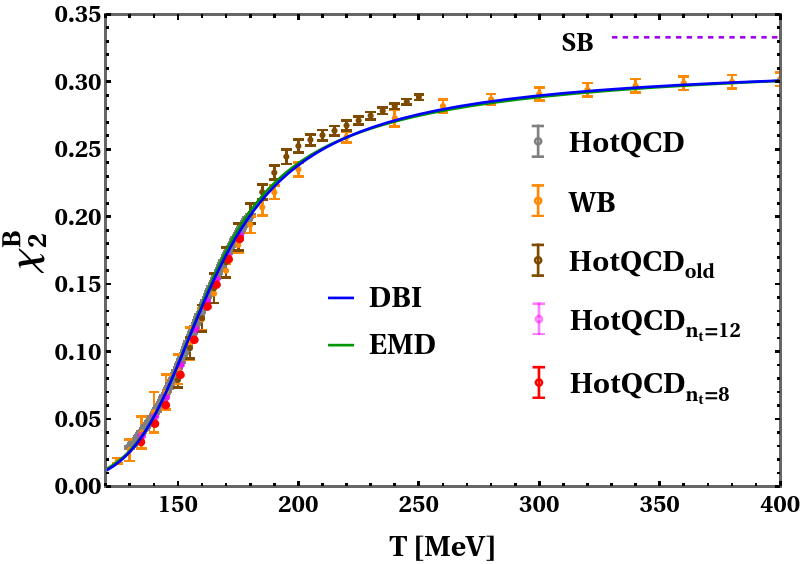}
  \caption{}
  \label{fig:chi2}
\end{subfigure}
\begin{subfigure}{.49\textwidth}
  \centering
  \includegraphics[width=\linewidth]{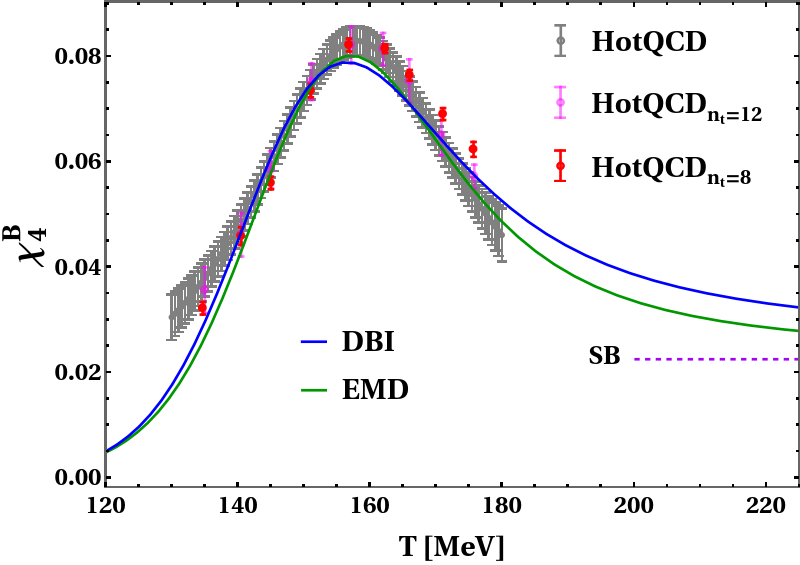}
  \caption{}
  \label{fig:chi4}
\end{subfigure}
\caption{Fits to the susceptibility data for $\chitwo$ (left panel) and $\chifour$ (right panel). The blue (green) curves correspond to the holographic fit of the DBI (EMD) model to the data. 
The Stefan--Boltzmann result for the susceptibilities of Appendix~\ref{app:inftemp} is also shown here.  The susceptibility data for $\chitwo$ from \cite{Borsanyi:2011sw} (orange) were used for the fit. The continuum extrapolated susceptibilities for $\chitwo$ and $\chifour$ of \cite{Bazavov:2020bjn} are shown in gray, which were used in the fits. Finite lattice size results of \cite{Bazavov:2020bjn} and the older data of \cite{HotQCD:2012fhj} are also displayed, but are not used in the fits.} 
\label{fig:cumulants}
\end{figure}

\paragraph{Susceptibilities:} 
The data for the susceptibilities fixes the potential $w(\phi)$ in the DBI model and the potential $Z(\phi)$ in the EMD model (see Appendix~\ref{app:emd}). Both susceptibilities $\chitwo$ and $\chifour$ are fit simultaneously. The goodness of the fit is measured by a similar statistical measure (\ref{eq:chisquared}) as what was used for the entropy fit, see below.
\begin{itemize}
    \item HotQCD: The data for $\chitwo$ and $\chifour$ that we use for the fit is the continuum data from  \cite{Bazavov:2020bjn}. We have chosen to fit the data from 130 to 180 MeV in steps of 5 MeV. 
    The continuum data is shown in Figs.~\ref{fig:chi2}~and~\ref{fig:chi4} in gray. 
    \item Wuppertal-Budapest: The data for $\chitwo$ we use for the fit is the continuum data from \cite{Borsanyi:2011sw}.  
    We will fit all the available data for $\chitwo$ up to 400 MeV (varying step size, but consistent with HotQCD data). 
    This data is shown in Fig.~\ref{fig:chi2} in orange.
\end{itemize}

Ideally, we would like to use the measure of~\eqref{eq:chisquared}, which properly takes into account the errors, also when fitting the susceptibility data. This however does not work nicely. The reason is that the recent HotQCD dataset for $\chitwo$ has way smaller errors than other data we are using, while it covers a rather narrow range of temperatures. Therefore direct fit to data would be totally dominated by this dataset, leading to a model that precisely describes the $\chitwo$ of QCD in the range of the dataset, but is likely a poor description for other observables or in other temperature ranges. This kind of fit is not the goal of this article.

In order to fix the issue, one could perhaps add extra parameters to our fit function so that the fit could become sensitive also to the other datasets. We however prefer to keep the fit function simple, so we change the fitting measure, so that it takes into account the other sets more efficiently. A possibility would be to use a similar measure as for the pure glue data~\eqref{eq:gluechi2}, which simply ignores all errors. We use instead a weighted version of~\eqref{eq:chisquared}:
\be\label{eq:chisquaredchis}
\chi^2_\mt{susc} = \frac{1}{n} \sum_{i=1}^n W_i \frac{(X_i-X^\mathrm{fit}_i)^2}{\Delta X_i^2} \ .
\ee
By choosing the coefficients $W_i$ (constant for each dataset) appropriately we can mitigate the issue, while keeping the temperature dependence of the error estimates. We use $W^{\mathrm{HotQCD}\chitwo}_i=1/100$ for the HotQCD $\chitwo$ data, $W^{\mathrm{WB}\chitwo}_i=1$ for the Wuppertal-Budapest $\chitwo$ data, and $W^{\mathrm{HotQCD}\chifour}_i=3$ for the HotQCD $\chifour$ data.

Our final fits are shown in Fig.~\ref{fig:cumulants}. Again, the  blue (green) curves are fits in the DBI (EMD) model. In the $\chitwo$ fit of Fig.~\ref{fig:chi2}, the two curves almost overlap. Notice that we only use the continuum HotQCD results (shown in gray) in the fit while we also show the results at finite lattice size (magenta and red data) in the plots. The fitted parameters of the  potential $w(\phi)$ in the DBI model and the potential $Z(\phi)$ in the EMD model are given in Table~\ref{tab:fitresultsDBI}~and~\ref{tab:fitresultsEMD}, respectively.

\begin{figure}[H]
\centering
\begin{subfigure}{.49\textwidth}
  \centering
  \includegraphics[width=\linewidth]{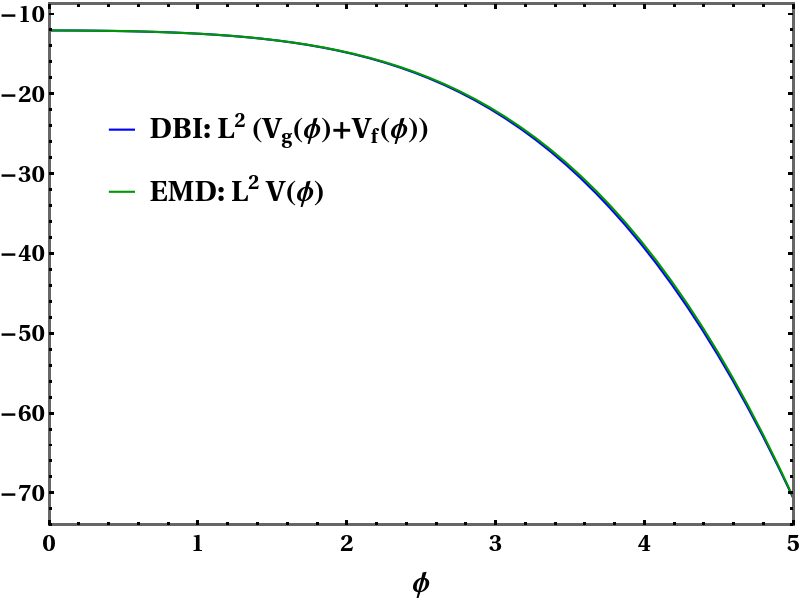}
  \caption{}
  \label{fig:Vpot}
\end{subfigure}
\begin{subfigure}{.49\textwidth}
  \centering
  \includegraphics[width=\linewidth]{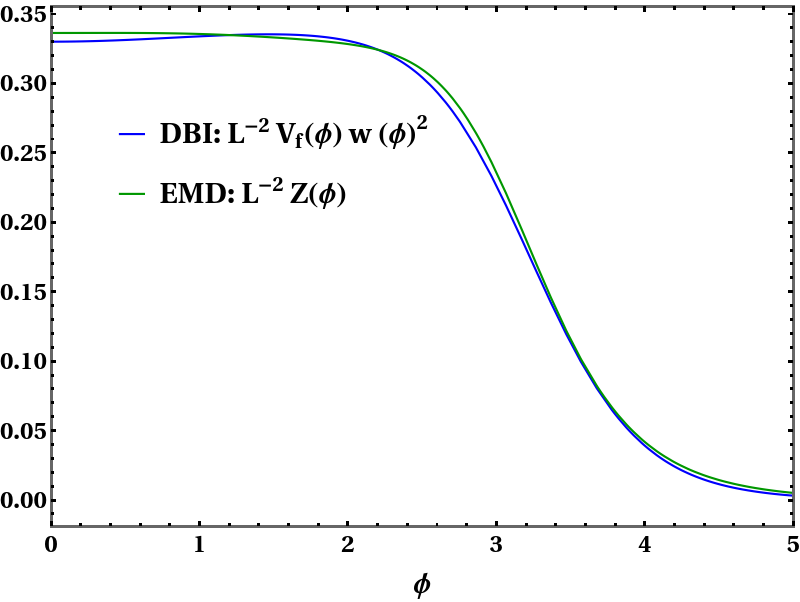}
  \caption{}
  \label{fig:fpot}
\end{subfigure}
\caption{The results for the fitted potentials. The left (right) panel shows the functions fitted to the entropy density (susceptibility) data. The blue (green) curves show the results in the DBI (EMD) model. Only the combinations $\Vg(\phi)+\Vf(\phi)$ and $\Vf(\phi)w(\phi)^2$ of DBI are shown as these match with the potentials $V(\phi)$ and $Z(\phi)$ of EMD at small densities. 
}
\label{fig:potentials}
\end{figure}

The fits shown in Figs.~\ref{fig:entropies}~and~\ref{fig:cumulants} correspond to the potentials shown in Fig.~\ref{fig:potentials}. The profiles of the potentials between the DBI and EMD setups are strikingly similar, when compared through the mapping~\eqref{eq:DBIEMDidentification}. The functions are multiplied by appropriate powers of the curvature scale $L$ to make them comparable. Some similarity is expected, as both models were fitted to the same lattice data which only probes the behavior of the equation of state at relatively low chemical potential. The agreement is however almost perfect given the fact that the parametrization of the fits to the susceptibility data are slightly different between the two models. As we pointed out above, our fits to the susceptibility data in Fig.~\ref{fig:potentials} are smooth near $\phi = 0$ and therefore missing the spiky feature appearing in several other fits in the literature~\cite{Knaute:2017opk,Critelli:2017oub,Cai:2022omk,Li:2023mpv,Hippert:2023bel,Fu:2024wkn}.

\begin{table}
\caption{Fitted parameters in the DBI model.}
\label{tab:fitresultsDBI} 
\[
\begin{array}{ |c|c|c|c|c|c|c|c|c|c|c| }
 \hline
 \text{Model} & \kappa_5^2 & \Lambda \ [\text{MeV}] & v_2 & v_4 & v_6 \\ 
 \hline
 \text{DBI} & 27.0239 & 1356.2478 & 1.4969 & -0.01994 & -0.001231 \\ 
 \hline
\end{array}
\]
\[
\begin{array}{ |c|c|c|c|c|c| }
 \hline
 \text{Model} & \gamma_1 & f_0 & f_2 & f_4 & f_6 \\ 
 \hline
 \text{DBI} & 0.5327 & 5.6879 & 0.04278 & 0.007703 & 0.002047 \\ 
 \hline
\end{array}
\]
\[
\begin{array}{ |c|c|c|c|c|c|c| } 
 \hline
 \text{Model} & \gamma_2 & w_0 & w_2 & w_4 & w_6 \\ 
 \hline
 \text{DBI} & 1.0873 & 3.0091 & -1.7916 & -0.1657 & -0.007432 \\ 
 \hline
\end{array}
\]
\end{table}

\begin{table}
\caption{Fitted parameters in the EMD model.}
\label{tab:fitresultsEMD} 
\[
\begin{array}{ |c|c|c|c|c|c|c| }
 \hline
 \text{Model} &  \kappa_5^2 & \Lambda \ [\text{MeV}] & \tilde{\gamma}_1 & b_2 & b_4 & b_6 \\ 
 \hline
 \text{EMD} & 10.2508 & 1075.7003 & 0.6082 & 1.9099 & -0.02862 & 0.001619 \\ 
 \hline
\end{array}
\]
\[
\begin{array}{ |c|c|c|c|c|c| }
 \hline
 \text{Model} & \tilde{\gamma}_2 & z_0 & z_2 & z_4 & z_6 \\ 
 \hline
 \text{EMD} & 1.2257 & 2.9700 & -2.2351 & -0.2654 & -0.01767 \\ 
 \hline
\end{array}
\]
\end{table}

\begin{table}
\caption{The values of the statistical measures and the sizes of the datasets for the fits.}
\label{tab:chisquared}
\[
\begin{array}{ |c|c|c|c| } 
 \hline
 \text{Model} & \chi^2_\text{glue}, n = 44 & \chi^2_\text{entropy}, n=48& \chi^2_\text{susc}, n=47 \\ 
 \hline
 \text{DBI} & 0.013 & 0.138 & 2.034 \\ 
 \hline
 \text{EMD} & - & 0.129 & 2.940 \\
 \hline
\end{array}
\]
\end{table}

\section{Results}\label{sec:results}

In this section we will discuss results for other 
observables than the entropy density and low-order susceptibilities which were used in the fits. 
We will divide this section in two parts, vanishing density followed by discussions at nonzero density relevant for quark-gluon plasma phase.

\subsection{Results at vanishing chemical potential}\label{sec:resultsmu0}

Let us begin the discussions at observables defined at vanishing baryon chemical potential, \ie, the zero density equation of state and the (higher-order) susceptibilities of the pressure around the $\muB=0$ result. This is an important cross-check of our results. 

\begin{figure}[hbt!]
\centering
\includegraphics[width=0.7\linewidth]{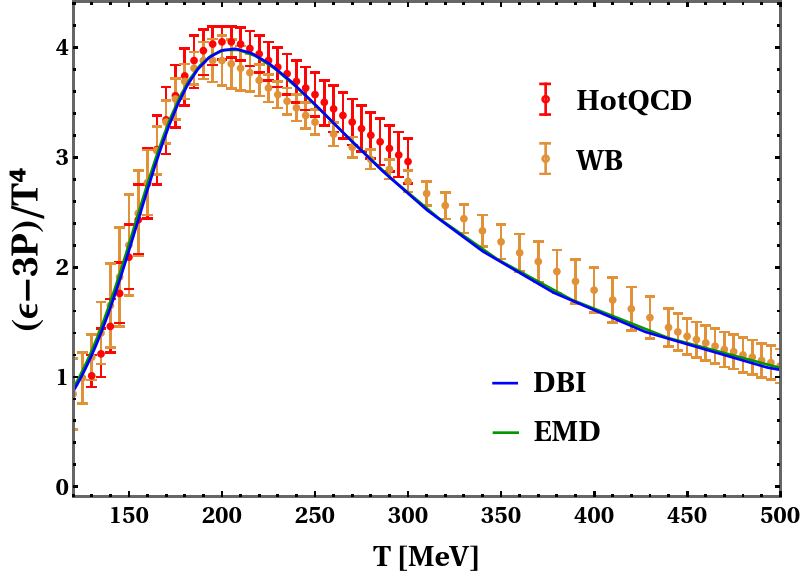}
  \caption{The interaction measure $\epsilon-3 P$ (\ref{eq:interactionmeasure}), normalized by $T^4$, at zero baryon chemical potential for the two fits is depicted as a function of $T$. It measures the conformality of the theory as the magnitude of the trace of the energy-momentum tensor. Predictions from both the DBI and the EMD models nicely match onto lattice results.
  The red data is from the HotQCD collaboration \cite{HotQCD:2014kol} and the orange data is from the Wuppertal--Budapest collaboration \cite{Borsanyi:2013bia}. 
  }
  \label{fig:intmea}
\end{figure}

Recall that our holographic models were fitted to specific gluon and lattice QCD results, so it is important that other available data are also well produced.  
We focus first on the interaction measure (\ref{eq:interactionmeasure}). In Fig.~\ref{fig:intmea}, we present the computation results from the holographic models alongside lattice data from \cite{Borsanyi:2013bia,HotQCD:2014kol}. Recall that the holographic analysis involves the 
integration in (\ref{eq:Pintegration}), where the contribution from very low temperatures to the EoS was assumed to be negligibly small. 
Remarkably, the predictions from the holographic models closely agree with the  
data, providing a satisfying consistency check. That is, the integration does not lead to any significant bias or offset between the predictions from the models and the lattice data. 

Let us then continue with the equation of state 
at nonzero but small chemical potential, characterized by the susceptibilities. Recall that we fitted the first two non-vanishing baryon number susceptibilities $\chitwo$ and $\chifour$ to the model parameters. The lattice QCD has results also for higher-order susceptibilities  $\chisix$ and $\chieight$ albeit with bigger error margins. Nevertheless, it is interesting to compare them with our results, and also to give predictions for even higher-order susceptibilities. 

\begin{figure}[hbt!]
\centering
\begin{subfigure}{.49\textwidth}
  \centering
  \includegraphics[width=\linewidth]{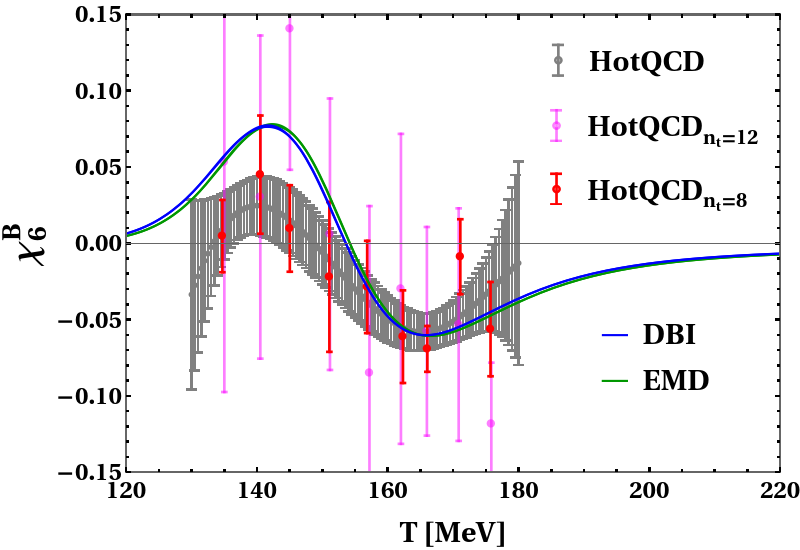}
  \caption{}
  \label{fig:chi6}
\end{subfigure}
\begin{subfigure}{.49\textwidth}
  \centering
  \includegraphics[width=\linewidth]{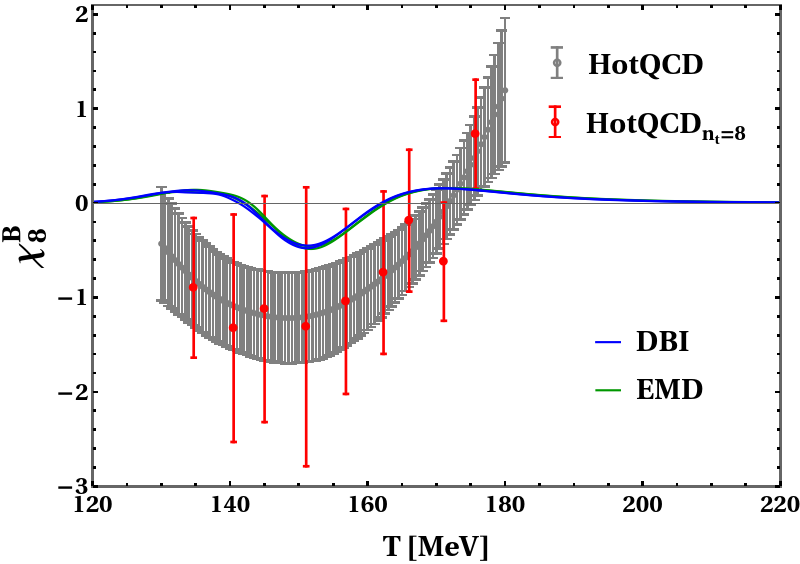}
  \caption{}
  \label{fig:chi8}
\end{subfigure}
\caption{{\bf{Left}}: Sixth order susceptibility $\chisix$ as a function of temperature at vanishing density. {\bf{Right}}: Eight order susceptibility $\chieight$ as a function of temperature at vanishing density. In both plots we display the results from both holographic models and the lattice data available from HotQCD collaboration \cite{Bazavov:2020bjn}. The gray band is the continuum extrapolated value. Finite size lattice results are also shown.}
\label{fig:chi6chi8}
\end{figure}

\begin{figure}[hbt!]
\centering
\begin{subfigure}{.49\textwidth}
  \centering
  \includegraphics[width=\linewidth]{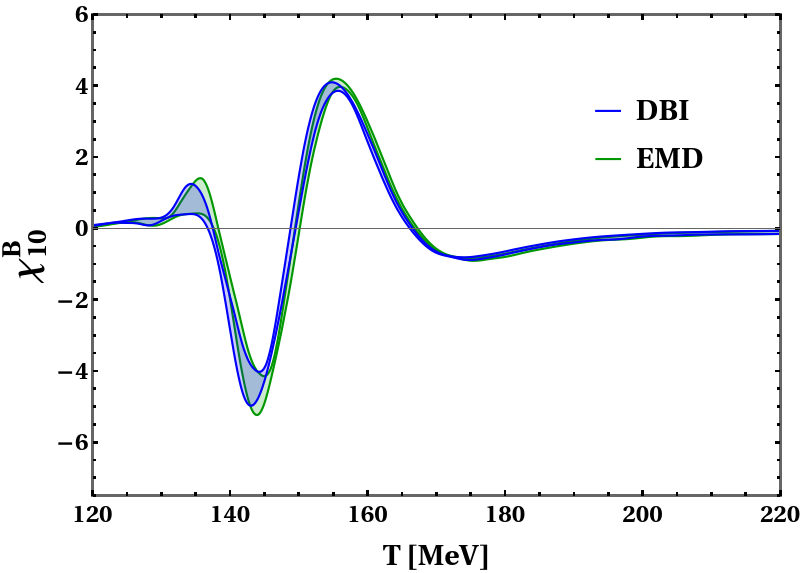}
  \caption{}
  \label{fig:chi10}
\end{subfigure}
\begin{subfigure}{.49\textwidth}
  \centering
  \includegraphics[width=\linewidth]{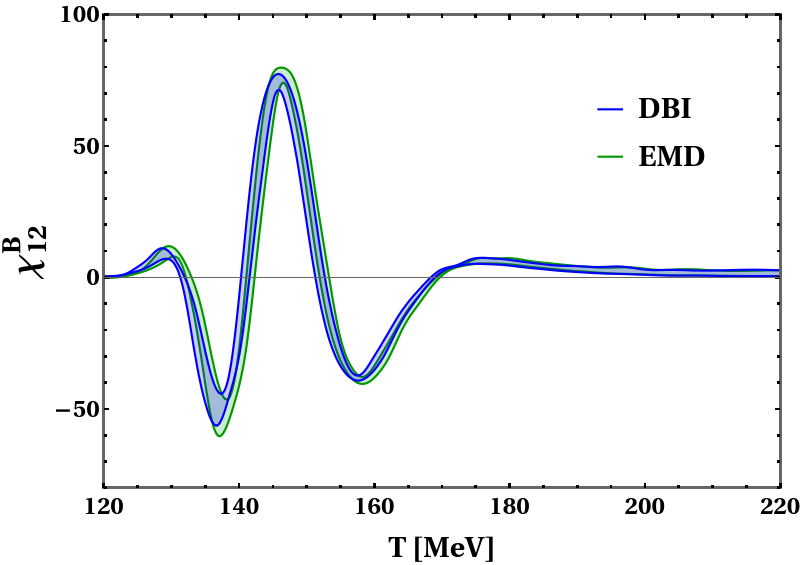}
  \caption{}
  \label{fig:chi12}
\end{subfigure}
\caption{{\bf{Left}}: Tenth order susceptibility $\chiten$ as a function of temperature at vanishing density. {\bf{Right}}: Twelfth order susceptibility $\chitwelve$ as a function of temperature at vanishing density.}
\label{fig:chi10chi12}
\end{figure}

To this end, the higher-order susceptibilities $\chisix$, $\chieight$, $\chiten$, and $\chitwelve$ were computed in the holographic models by fitting a 13th order polynomial in chemical potential to the density data. We generated density data at fixed temperature up to a limiting value of the chemical potential $\muB^{\text{max}}/T$, which  we varied in the set $\muB^{\text{max}}/T \in \{1.85, 2.00, 2.15\}$ to estimate numerical accuracy. The range was chosen such that the fit results show maximal stability. The resulting variation in the fit results was used to obtain the bands shown in Figs. \ref{fig:chi6chi8} and \ref{fig:chi10chi12}. The bands in Fig.~\ref{fig:chi6} are barely visible because they are rather narrow.  
Notice that the chosen values for $\muB^{\text{max}}$ 
were chosen to  
depend linearly on the temperature. The reason for this is that the natural expansion parameter of dimensionless susceptibilities in (\ref{eq:pressure}) is $\muB/T$. By using fixed $\muB^{\text{max}}$ at all temperatures would lead to poor precision either at low or high temperature end of the data.

We compare our results to the lattice data~\cite{Bazavov:2020bjn} for the susceptibilities $\chisix$ and $\chieight$ in Fig.~\ref{fig:chi6chi8}. Our results are similar to earlier studies~\cite{Li:2023mpv}, and show agreement with the temperature dependence observed on the lattice for $\chisix$, whereas for $\chieight$ the situation is less clear as the error margins in the lattice data are also larger. Moreover, we observe that the results from the DBI and EMD models are virtually identical. That is, the fit to $\chitwo$ and $\chifour$ appears to fix the results also for the higher-order susceptibilities regardless of the choice of the flavor action, \ie, the action for the five-dimensional gauge fields. This happens even though the integral formulas~\eqref{eq:Idef} and~\eqref{eq:munBrelEMD} for the DBI and EMD actions, respectively, which determine the chemical potentials as a function of the baryon number and therefore the susceptibilities, have different functional forms in the two models. This implies that the differences in the backgrounds at nonzero baryon number must compensate for the difference between the integral formulas.

We also show predictions for even higher-order susceptibilities $\chiten$ and $\chitwelve$ in Fig.~\ref{fig:chi10chi12}. The predictions from the EMD and DBI models are still similar, even though the difference between the models slightly grows with increasing order of the susceptibility. Therefore it is challenging to distinguish the two models just based on susceptibility data.

\subsection{Results at nonzero chemical potential}\label{sec:resultsmune0}

In this subsection we use the holographic models fitted to lattice results as described earlier to gain insight on various quantities at non-vanishing density. While the holographic models can be easily extended all the way to the high-density neutron-star matter regime, in this paper we are content with working in moderate densities. 
    
\begin{figure}[hbt!]
\centering
\includegraphics[width=0.7\linewidth]{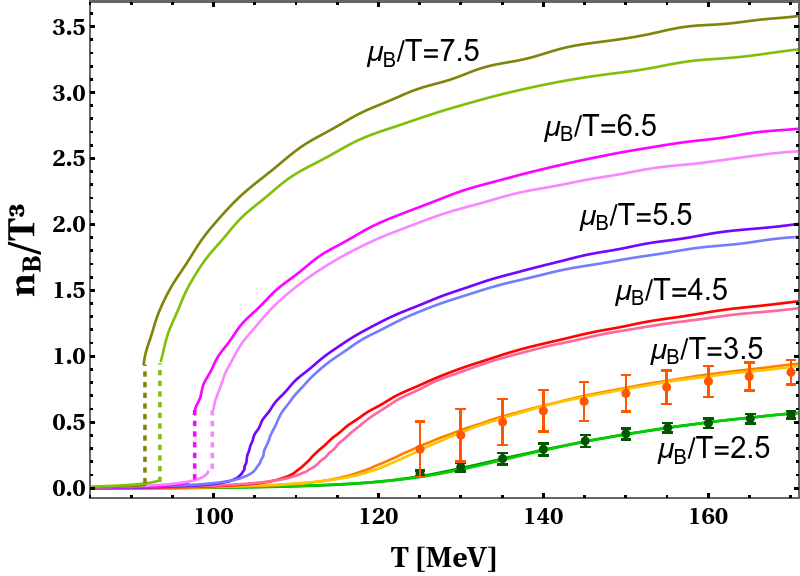}
  \caption{We show the baryon number density $n_\mathrm{B}$ for various values of $\mu_\mathrm{B}/T$. The results for both DBI and EMD models are shown. The results for DBI (EMD) correspond to the curves that are higher (lower) and darker (lighter) in color. The lattice data is from a simulation with 2+1+1 flavors \cite{Borsanyi:2021sxv}. The critical point is near $\mu_\mathrm{B}/T \approx 6$ both in the EMD and DBI models. For higher values of $\mu_\mathrm{B}/T$ we have a first order phase transition which is signaled by a discontinuity in the density. This discontinuity is shown in the figure by dashed lines.}
  \label{fig:finitedensity}
\end{figure}

In Fig.~\ref{fig:finitedensity} we show our results for the baryon number densities at nonzero density for both DBI and EMD models. It matches well with the lattice data~\cite{Borsanyi:2018grb} that is available at low densities. Notice that this lattice data is from a simulation including the charm quark whereas our model was fitted to data from simulations without the charm quark. The agreement is good nevertheless.
There is a critical end point in both models near $\mu_\mathrm{B}/T \approx 6$ and a first order transition for 
larger values of $\mu_\mathrm{B}/T$. 
The location of the CEP is in the same ballbark as in other holographic works~\cite{DeWolfe:2010he,DeWolfe:2011ts,Knaute:2017opk,Demircik:2021zll,Grefa:2021qvt,Li:2023mpv,Hippert:2023bel,Chen:2024ckb,Fu:2024wkn}. Interestingly, our value is also close to the numbers obtained by using the functional renormalization group method~\cite{Fu:2019hdw,Gao:2020qsj}. 

\begin{figure}[hbt!]
\centering
\includegraphics[width=0.7\linewidth]{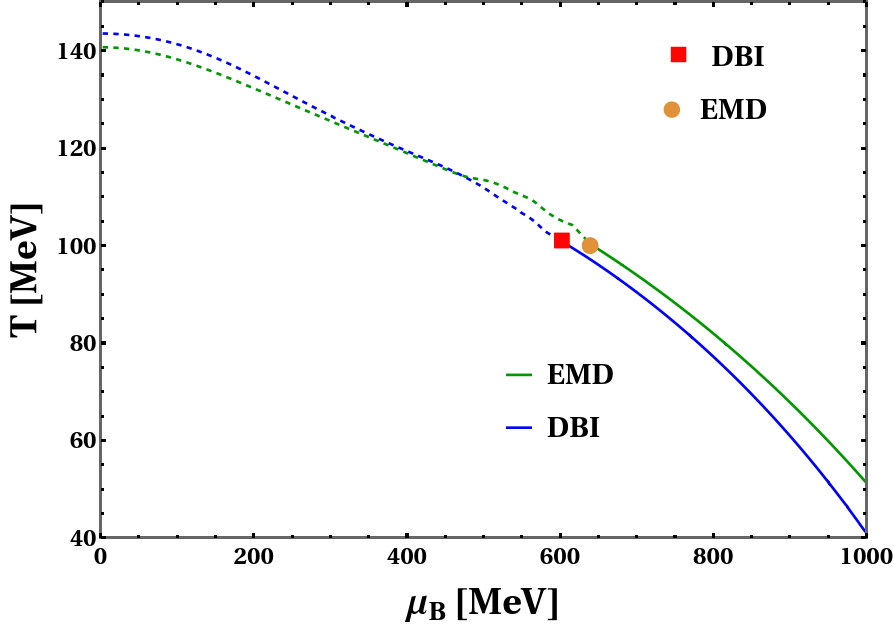}
\caption{We depict the phase diagram in the holographic model as inferred from the fitted potentials. The dashed curves trace the minimum of the adiabatic speed of sound \eqref{eq:c2isentropic} and can be thought of as a proxy for the crossover transition line.  The solid curves show the line of first order phase transitions. The two types of curves are connected at critical points, marked with a square (DBI) and a disk (EMD) for which the critical values read  $(\muB,T) = (602.5\ \text{MeV},100.9\ \text{MeV})$ and $(\muB,T) = (639.3\ \text{MeV},100.4\ \text{MeV})$, respectively.}
\label{fig:cp}
\end{figure}

The phase diagram and the location of the critical end point is shown explicitly in Fig.~\ref{fig:cp}. 
The first order transition line (solid curves) was computed by using the Maxwell construction at constant chemical potential. We have checked that the Maxwell construction along constant temperature slices gives identical results. 
At low densities, there is a single saddle in the gravity path integral, and therefore no phase transition.  But as a proxy for the crossover transition, or a Widom line \cite{Sordi:2023cjq}, we track the position of the minimum of the isentropic speed of sound:
\begin{equation}\label{eq:c2isentropic}
    c^2_{s/\nB} = \bigg(\frac{\partial P}{\partial \epsilon}\bigg)_{s/\nB}  = \frac{\nB^2 \frac{c_V}{T}+s^2T^2\chitwo-2 \nB s \chi_{\muB T}}{\left(Ts+\muB \nB\right)\left(Tc_V\chitwo - \chi_{\muB T}^2\right)}\ ,
\end{equation}
where we have introduced the heat capacity $c_V=T(\partial s/\partial T)$ and $\partial_T \nB=\partial_{\muB} s=\chi_{\muB T}$.
Note that the first order transition occurs between two black hole phases in the holographic models. Therefore, both phases are deconfining, and neither of the models have a proper deconfinement phase transition. The confining phase has not been captured in a satisfactory fashion.

\begin{figure}[hbt!]
\centering
\includegraphics[width=0.5\linewidth]{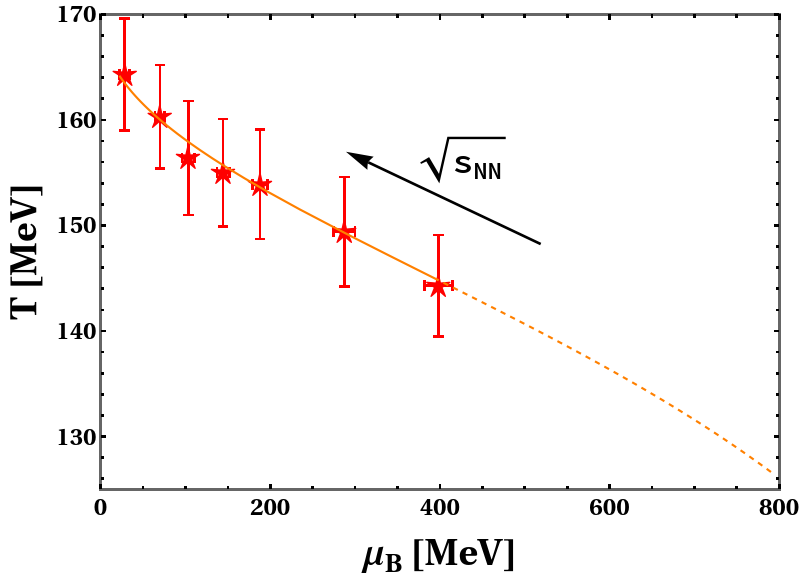}
  \caption{The chemical freeze-out parameters for the events $\sqrt{\sNN} \in \{7.7,11.5,19.6,27,39,62.4,200\}$ GeV in the $(\muB,T)$-plane. The data is from \cite{STAR:2017sal} Table VIII for 0-5~\% centrality. The orange line is the fitted curve (\ref{eq:freeze-outcurve}). Dashed curve denotes extrapolation of the data.}
  \label{fig:diagram}
\end{figure}

We will now compare our results with data from heavy-ion collisions, particularly from the BES program at RHIC. In high-energy heavy-ion collisions, the key controllable variables are the colliding ions, the centrality of the collision, and the collision energy.
Observables are often listed as a function of the collision energy $\sqrt{\sNN}$ in the center-of-mass frame. In our holographic models we instead characterize a given state of the system by temperature and chemical potential. Thus we would want to have a way to map a given point in the $(\muB,T)$-plane to some value of the collision energy $\sqrt{\sNN}$. The collision data allows one to infer this at freeze-out. We use the freeze-out parameters given in Table VIII of \cite{STAR:2017sal}. These parameters are shown in Fig.~\ref{fig:diagram} for 0-5~\% centrality. 

The collision data allows us to determine the freeze-out parameters for fixed values of $\sqrt{\sNN}$. In Fig.~\ref{fig:diagram} these are given for $\sqrt{\sNN} \in \{7.7,11.5,19.6,27,39,62.4,200\}$ GeV. To make predictions outside of the range of these center-of-mass energies we fit the following curves
\be
\label{eq:freeze-outcurve}
    T_{\mathrm{CF}}(\sqrt{\sNN}) = \frac{a_1}{1+\exp\left(b_1-\log(\sqrt{\sNN})/c_1\right)} \ , \quad \mu_{\mathrm{CF}}(\sqrt{\sNN})= \frac{a_2}{1+b_2\sqrt{\sNN}}
\ee
to the shown data (CF refers to chemical freeze-out). The fitted parameter values are given in Table~\ref{tab:freeze-outparams}.
\begin{table}[!ht]
\caption{The parameter values for the freeze-out curves.}
\label{tab:freeze-outparams}
\[
\begin{array}{ |c|c|c|c|c| } 
 \hline
 a_1 \ [\text{MeV}] & b_1 & c_1  & a_2 \ [\text{MeV}] & b_2  \\ 
 \hline
169.0 & -0.733 & 1.950 & 1210.0& 0.267\\ 
 \hline
\end{array}
\]
\end{table}

Next, we connect our results with the BES collision data from the STAR collaboration \cite{STAR:2021iop}.
We are interested in the higher moments of the distributions of conserved quantities $(N)$
because they are more sensitive to the finite correlation length \cite{Stephanov:2011pb,STAR:2020tga}. These can be analyzed for any conserved quantity but we are only interested in the case where $N$ is the net-proton number in the collision.
Of particular relevance are the skewness $S=\langle (\delta N)^3\rangle/\sigma^3$ and the kurtosis $\kappa=[\langle (\delta N)^4\rangle/\sigma^4]-3$ of the distribution of the  net-proton number fluctuations $\delta N=N-M$,  where $M=\langle N\rangle$ is the mean and $\sigma$ is the standard deviation. To map the nomenclature between the susceptibilities $\chin$ computed here and the net-proton cumulants $C_i$ in \cite{STAR:2021iop} for which $C_1=M$, $C_2=\sigma^2$, $C_3=S\sigma^3$, and $C_4=\kappa\sigma^4$ one can show 
that~\cite{Gupta:2011wh} 
\be
C_n = V_\mathrm{fr}T^3\chin \ ,
\ee
where $V_\mathrm{fr}$ is the (average) freeze-out volume of the quark-gluon plasma produced in the collision. Notice that the definition of this volume may contain significant uncertainties, but we will only consider ratios between different $C_n$, where the volume factor cancels.
The data they provide are for the ratios of $C_i$ and so conveniently $C_2/C_1=\chitwo/\chione$, $C_3/C_2=S\sigma=\chithree/\chitwo$, and $C_4/C_2=\kappa\sigma^2=\chifour/\chitwo$. Notice that $\chione = \nB/T^3$ and $M = V_\mathrm{fr} \nB$.

\begin{figure}[hbt!]
\centering
\begin{subfigure}{.49\textwidth}
  \centering
  \includegraphics[width=\linewidth]{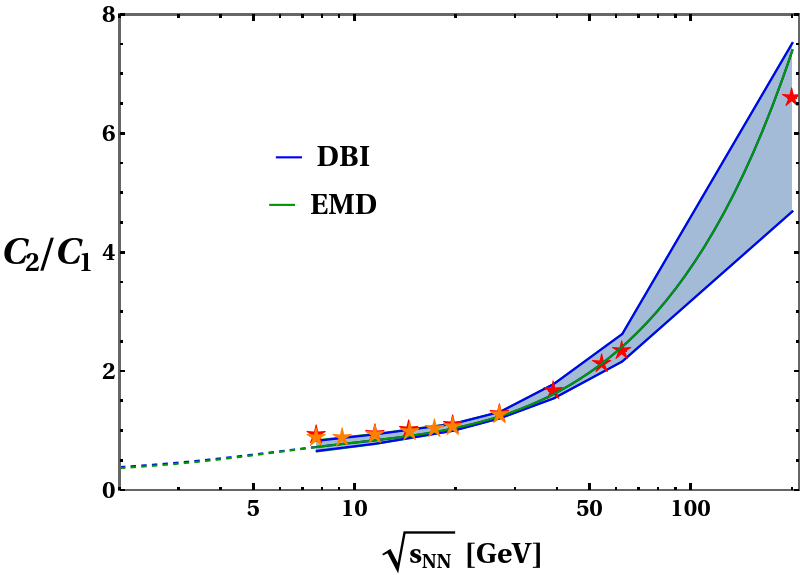}
  \caption{}
  \label{fig:C2C1}
\end{subfigure}
\begin{subfigure}{.49\textwidth}
  \centering
  \includegraphics[width=\linewidth]{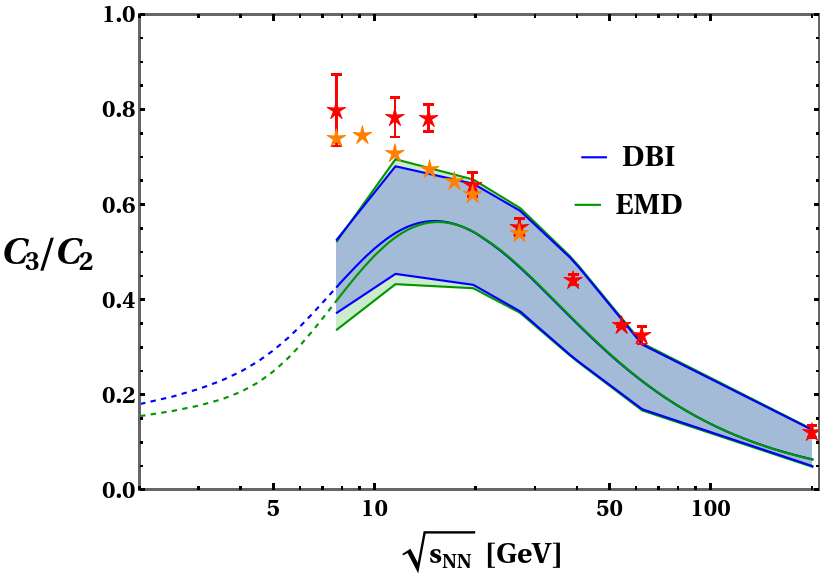}
  \caption{}
  \label{fig:C3C2}
\end{subfigure}
\begin{subfigure}{.49\textwidth}
  \centering
  \includegraphics[width=\linewidth]{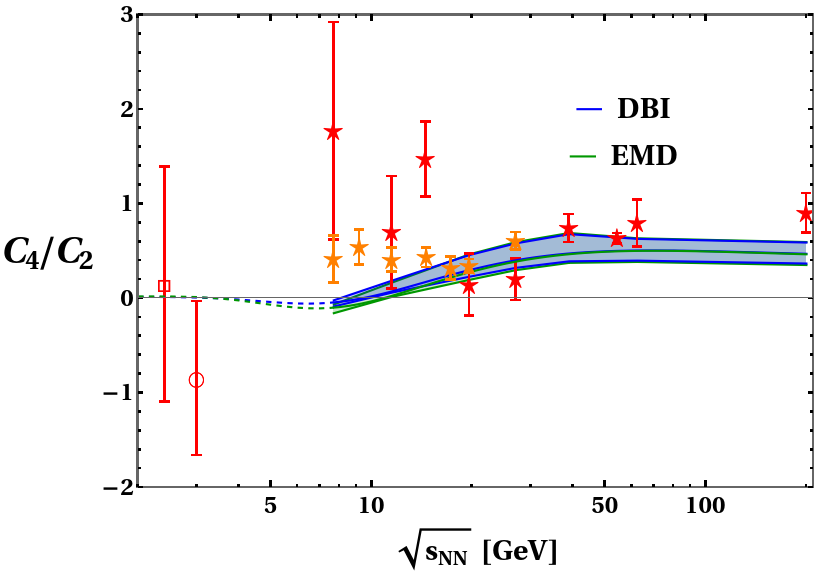}
  \caption{}
  \label{fig:C4C2}
\end{subfigure}
\caption{Ratios of cumulants as a function of the collision energy $\sqrt{\sNN}$. The results for both DBI and EMD models are shown in the figures. The collision data is from \cite{STAR:2021iop} for Au+Au 0-5 \% centrality. The different collision energies are $\sqrt{\sNN}\in\{7.7,11.5,14.5,19.6,27,39,54.4,62.4,200\}$ GeV. We have also included the recent preliminary data from BES-II~\cite{Pandav:CPOD2024} in the plots for comparison (orange stars), but recall that they are not contributing to the fit of the freeze-out curve shown in Fig.~\ref{fig:diagram}.
In the bottom plot we have also included the recent data point from STAR \cite{STAR:2021fge} at $\sqrt{\sNN}=3.0$ GeV (open disk) and the $\sqrt{\sNN}=2.4$ GeV data point (open square) from  0\%–10\% collisions obtained by the HADES collaboration \cite{HADES:2020wpc}.
It should be emphasized that these results are predictions of our holographic models; no fitting to data was performed.
}
\label{fig:star}
\end{figure}

The results for the ratios $C_2/C_1$, $C_3/C_2$, and $C_4/C_2$ are shown in Figs.~\ref{fig:C2C1},~\ref{fig:C3C2},~and~\ref{fig:C4C2}, respectively. STAR data from BES-I~\cite{STAR:2021iop}  (BES-II~\cite{Pandav:CPOD2024}) is shown as the red (orange) stars and error bars. We also show the recent data points from STAR~\cite{STAR:2021fge} at $\sqrt{\sNN}=3.0$ GeV and from HADES~\cite{HADES:2020wpc} at $\sqrt{\sNN}=2.4$ GeV for $C_4/C_2$.
In these plots, we depict curves that correspond to the freeze-out curve as defined in (\ref{eq:freeze-outcurve}), \ie, we have computed the cumulants in the holographic model along this freeze-out curve. The curves in the range of the data are shown as solid curves and their extrapolations to lower $\sqrt{\sNN}$ are shown as dashed curves. The depiction also includes bands, which illustrate the experimental uncertainty of the chemical freeze-out curve, as shown in Fig.~\ref{fig:diagram}, and the consequent 
uncertainty in the mapping between $(\muB,T)$ and $\sqrt{\sNN}$. For each point $\sqrt{\sNN}$, we determine the range of possible values for the cumulant ratio within the rectangle outlined in Fig.~\ref{fig:diagram} and subsequently interpolate to identify the maximum extents of these ranges.
We observe that the agreement between the holographic model and the STAR data is good in general, but there is a tendency that the model data for the ratio $C_3/C_2$ is systematically below the experimental data. This tension was also observed in~\cite{Li:2023mpv}, where it was compensated by adjusting the fit to the freeze-out curve.  
Indeed, note that $C_2 \propto \chitwo$ depends strongly on the temperature near freeze-out. This leads to a rather large error estimate (\ie, wide band in the plot) for $C_3/C_2$.

Note that the recent BES-II data, as shown in Fig.~\ref{fig:star}, seems to point to smaller $C_4/C_2$ at higher density than the run BES-I. This is also preferred by our model, while some tension between the data and the model remains. 
We find that $C_4/C_2$ becomes suppressed at higher densities (smaller  $\sqrt{\sNN}$), and perhaps even becomes negative in our model. The sign change of this quantity could be linked with the existence of the CEP \cite{Stephanov:2011pb}. 
It is important to note though that the CEP as inferred from our holographic model is still far from the range of the experimental energy scan.

\section{Discussion and outlook} \label{sec:discussion} 

Holographic QCD has reached the level of sophistication that allows for a detailed reproduction of numerous lattice QCD outcomes, a development we have thoroughly explored within this study. We demonstrated that fitting the model systematically to data for the entropy density and the lowest susceptibilities $\chitwo$ and $\chifour$ fixes the predictions relatively tightly. In particular, we obtained almost identical results for the equation of state and the phase diagram at nonzero baryon number density in the two classes of models we considered, \ie, the DBI and EMD models. Building on this, we envisage a similar systematic analysis in two-flavor holographic QCD models~\cite{Zhao:2023gur,Liu:2023pbt}.

The framework is not perfect though and falls short in some obvious aspects. With the family of models discussed in this work, the most striking shortcoming is the complete ignorance of the confining phase. In addition, the current discussion was limited to the description of chirally symmetric phase with massless flavors and hence the effects due to the mass of the strange quark were not implemented. Moreover, these models do not include nuclear matter, which may already have a significant effect on the equation of state at the chemical potentials where the critical points are found. Therefore the predictions for the phase diagram and in particular the suggested location of the critical point should be taken with a grain of salt. It is likely that the uncertainties due to these missing features are significantly larger than the uncertainties due to the fitting procedure.
Recent works incorporating flavor-dependent chemical potentials in a holographic setting include \cite{Jarvinen:2023xrx,Kovensky:2023mye}, where $\mu_\text{isospin}\ne 0$,  and \cite{Jarvinen:2015ofa,CruzRojas:2024etx}, where $m_\text{strange\ quark}\ne 0$.  In the latter case, the anchoring of the model to the lattice results is lacking. For an attempt to include confinement and nuclear matter, see~\cite{Demircik:2021zll}, which however still assumes zero quark masses. We plan to address these shortcomings in future work.

One of the advances in using the holographic approach is that one can also obtain out-of-equilibrium results. For example, it would be interesting to compute the shear as well as the purely QCD contribution to bulk viscosity and the electrical and thermal conductivities as evaluated only recently in two holographic models \cite{Hoyos:2020hmq,Hoyos:2021njg}. Extracting these quantities would be straightforward, since the master formulas given in \cite{Hoyos:2020hmq,Hoyos:2021njg} are directly applicable to all actions written in the forms in eqns.~(\ref{eq:actionDBI})~and~(\ref{eq:actionEMD}) and simply require on-shell values of the bulk fields and their derivatives at the black hole horizon.

By construction, the DBI and EMD models as explored here match at vanishing baryon number density, and produce surprisingly similar results up to $\muB/T \approx 8$, as we have seen in this article.
The models may however function differently 
deep in the quark matter phase. It would be interesting to extend the current setup to neutron-star matter densities (see reviews~\cite{Jarvinen:2021jbd,Hoyos:2021uff} for holographic approach to neutron stars) and explore astrophysical constraints on the models, especially those set by the  
NICER results \cite{Jokela:2021vwy}. However, as will be discussed elsewhere \cite{CruzRojas:2024igr,Demircik:2024aig}, an instability that can be connected with the flavor anomaly of QCD appears to emerge already at moderate densities in the holographic models describing deconfined quark matter. The instability region is expansive but may be strongly dependent on the mass of the strange quark. Therefore, a pressing issue is to invest efforts in incorporating flavor-dependent description of the holographic matter and at bare minimum draw the boundaries of where the homogeneous ground state is to be trusted. Constructing the inhomogeneous ground state, especially to non-polynomial actions such as (\ref{eq:actionDBI}), is challenging, but not unprecedented \cite{Jokela:2014dba,Mukhopadhyay:2020tky,Ishigaki:2024kjz}.


\paragraph{Acknowledgments}
We would like to thank Oscar Henriksson and Carlos Hoyos for discussions. We thank Jere Remes for collaboration at the initial stages of this work. N.~J. has been supported in part by Research Council of Finland grants no. 345070 and 354533. The work of M.~J. was supported in part by an appointment to the JRG Program at the APCTP through the Science and Technology Promotion Fund and Lottery Fund of the Korean Government. M.~J. was also supported by the Korean Local Governments -- Gyeongsangbuk-do Province and Pohang City, and by the National Research Foundation of Korea (NRF) funded by the Korean government (MSIT) (grant number 2021R1A2C1010834). A.~P. acknowledges the support from the Jenny and Antti Wihuri Foundation.

\appendix

\section{Details of the background} \label{app:backgroundEOM} 
In this appendix we flesh out some necessary details of the background. We will divide the appendix in two parts, first dealing with the DBI case, and then discussing how the formulas alter when focusing on the EMD action.

\subsection{Dirac--Born--Infeld model}

The DBI action (\ref{eq:actionDBI}) yields the following equations of motion for this Ansatz:
\bea
    h \phi'' + \left(h'+h(4A'-B')\right)\phi' - e^{2B}\bigg( \Vg'(\phi)& & \nonumber\\
   +\Vf'(\phi)\sqrt{1-w(\phi)^2e^{-2(A+B)}\psi'^2} -w'(\phi) \frac{\Vf(\phi)w(\phi)e^{-2(A+B)}\psi'^2}{\sqrt{1-w(\phi)^2e^{-2(A+B)}\psi'^2}}\bigg)  & = & 0 \label{eq:eomdbiphi} \\
\psi'' + \bigg(2A'-B' + \left(\log\left(\Vf(\phi)w(\phi)^2\right)\right)'\phi' \bigg)\psi' & & \nonumber  \\
-w(\phi)^2 e^{-2(A+B)}\bigg(3A' + \left(\log\left(\Vf(\phi)w(\phi)\right)\right)'\phi' \bigg)\psi'^3 & = & 0 \label{eq:eomdbipsi} \\
A'' -A'B' + \frac{1}{6}\phi'^2 & = & 0 \label{eq:eomdbiA} \\
h'' + (4A'-B')h'-\frac{\Vf(\phi)w(\phi)^2e^{-2A}\psi'^2}{\sqrt{1-w(\phi)^2e^{-2(A+B)}\psi'^2}} & = & 0 \label{eq:eomdbih} \\
    h(24A'^2-\phi'^2) + 6A'h' & & \nonumber \\
    + 2e^{2B}\bigg(\Vg(\phi) + \Vf(\phi)\sqrt{1-w(\phi)^2e^{-2(A+B)}\psi'^2} +\frac{\Vf(\phi)w(\phi)^2 e^{-2(A+B)}\psi'^2}{\sqrt{1-w(\phi)^2e^{-2(A+B)}\psi'^2}}\bigg) & = & 0 \ .\label{eq:eomdbiB}
\eea
The last equation (\ref{eq:eomdbiB}) is a constraint. These equations of motion have trivial symmetries
\begin{align}
    &r \rightarrow r + r_0 \nonumber \nonumber \\
    &r \rightarrow \Lambda_h r\ , \qquad h \rightarrow \Lambda_h^2 h\ , \qquad \psi \rightarrow \Lambda_h \psi \label{eq:symmetries} \\ 
    & A \rightarrow A + \log \Lambda_A , \qquad \psi \to \Lambda_A \psi \nonumber \ .
\end{align}
We use the symmetries (\ref{eq:symmetries}) when solving for the fluctuations in Appendix~\ref{app:fluctuations}.
By integrating equation (\ref{eq:eomdbipsi}) we obtain the conserved charge 
\be\label{eq:charge1}
    \nh = \frac{\Vf(\phi)w(\phi)^2e^{2A-B}\psi'}{\sqrt{1-w(\phi)^2e^{-2(A+B)}\psi'^2}} \ ,
\ee
which as the notation suggests will be simply related to the baryon number density.
Another conserved charge can be obtained from integrating (\ref{eq:eomdbih})
\be\label{eq:charge2}
    H = h'e^{4A-B}-\nh \psi \ .
\ee
We work in the gauge $B=0$ from now on.

The asymptotically AdS geometry is defined by the following asymptotics
\bea
    A(r) & = & r/L + \mathcal{O}(e^{-2\nu r/L}) \label{eq:AUV}\\
    h(r) & = & 1 + \mathcal{O}(e^{-4 r/L})\label{eq:hUVbc} \\
    \phi(r) & = & \phi_a e^{-\nu r/L} + \mathcal{O}(e^{-2\nu r/L}) \\
    \psi(r) & = & \psi_0^{\mathrm{far}} + \psi_2^{\mathrm{far}}e^{-2r/L} +\mathcal{O}(e^{-(2+\nu)r/L}) \ ,\label{eq:asympstand}
\eea
where $\nu = 4-\Delta$ and $\Delta = 2+\sqrt{4+m^2 L^2}$ is the scaling dimension of the operator dual to the dilaton with the dilaton UV mass $m$ defined in~\eqref{eq:UVpot}, and the radius of curvature is given by
\be
 L = \frac{L_0}{\sqrt{1-\Vf(0)L_0^2/12}} \ ,
\ee
with the radius $L_0$ defined in~\eqref{eq:Vgpot}.
The source parameter $\phi_a$ in the dilaton asymptotics in~\eqref{eq:asympstand} maps to the gauge coupling in field theory, and defines the overall energy scale of the theory through the dimensionful variable
\be \label{eq:Lambdadefapp}
 \Lambda = \phi_a^{1/\nu} \ .
\ee
This source parameter needs to be fixed for all solutions in the same theory, \ie, independently of temperature and chemical potential.

The thermodynamics are given by the standard gauge/gravity dictionary. The temperature of the dual theory is given by the Hawking temperature of the black hole 
\be
    T = \frac{e^{A(\rH)}}{4\pi}h'(\rH)\ ,
\ee
where $\rH$ is the location of the horizon. The entropy density is given by the Bekenstein--Hawking entropy
\be
    s = \frac{A_H}{4 G_5 V} = \frac{2\pi}{\kappa_5^2}e^{3A(\rH)} \ ,
\ee
where $V$ is the spatial volume. The baryon chemical potential is identified with the asymptotic value of the gauge field
\be
    \muB = \lim_{r\rightarrow \infty} \psi(r) = \psi_0^{\mathrm{far}}\ .
\ee
Integrating~\eqref{eq:charge1} gives, in the $B=0$ gauge,
\be \label{eq:munBrel}
 \muB =  \int_{\rH}^\infty \dd r\ \frac{\nh}{e^{2A}\Vf(\phi)w(\phi)^2}\left(1+\frac{\nh^2}{e^{6A}\Vf(\phi)^2w(\phi)^2}\right)^{-1/2}\ .
\ee
The baryon density is identified with the boundary value of the radial momentum conjugate to the gauge field 
\be
    \nB = \lim_{r\rightarrow \infty} \frac{\partial \mathcal{L}}{\partial (\partial_{r} \psi)} = \frac{\nh}{2\kappa_5^2} = -\frac{\Vf(0)w(0)^2\psi_2^{\mathrm{far}}}{\kappa_5^2 L} 
\ee
with the understanding $S=\int \dd^5x\mathcal{L}$.

\subsection{Einstein--Maxwell-dilaton model} \label{app:emd} 

The equations of motion read
\bea
h \phi'' + \left(h'+h(4A'-B')\right)\phi' -e^{2B}\bigg( V'(\phi)-\frac{e^{-2(A+B)}\psi'^2}{2} Z'(\phi)\bigg) & = & 0 \\
\psi'' + \bigg(2A'-B' + \left(\log{Z(\phi)} \right)'\phi' \bigg)\psi' & = & 0 \\
A'' -A'B' + \frac{1}{6}\phi'^2 & = & 0 \\
h'' + (4A'-B')h'-e^{-2A}Z(\phi)\psi'^2 & = & 0 \\
    h(24A'^2-\phi'^2) + 6A'h' +2e^{2B}V(\phi)+e^{-2A}Z(\phi)\psi'^2  & = & 0 \ .
\eea
This model has the conserved charges 
\bea
    \nh & = &Z(\phi)e^{2A-B}\psi' \\
    H & = & e^{2A-B}\big(h'e^{2A}-Z(\phi)\psi\psi'\big) \ .
\eea
As with the DBI action we continue working in the gauge $B=0$. Integrating the first of these equations, we obtain
\be \label{eq:munBrelEMD}
 \muB =  \int_{\rH}^\infty \dd r\ \frac{\nh}{e^{2A}Z(\phi)}\ .
\ee

This EMD model can be obtained from the DBI model by expanding the action to quadratic order in $\psi$. We see that in this limit the action (\ref{eq:actionEMD}) is obtained provided we identify $V(\phi) = \Vg(\phi)+\Vf(\phi)$ and $Z(\phi) = \Vf(\phi)w(\phi)^2$. By performing the same calculation for $\chifour$ in this EMD model as is done in Appendix~\ref{app:cumulants} for the DBI model, we see that 
\be
 \chi_2^{\mt{B},\mathrm{EMD}}(T) = \chi_2^{\mt{B},\mathrm{DBI}}(T) 
\ee
and
\be
\chi_4^{\mt{B},\mathrm{EMD}}(T) = \chi_4^{\mt{B},\mathrm{DBI}}(T) -\frac{3}{2\kappa_5^2}\frac{\int_{\rH}^\infty \dd r e^{-8A}\Vf(\phi)^{-3} w(\phi)^{-4}}{\left(\int_{\rH}^\infty \dd r e^{-2A}(\Vf(\phi)w(\phi)^2)^{-1}\right)^4}
\ee
if one would make the identification (\ref{eq:DBIEMDidentification}).
Notice, that this identification holds approximately between the DBI (\ref{eq:Vgpot})--(\ref{eq:wpot}) and EMD potentials (\ref{eq:Vpot})--(\ref{eq:Zpot}) we consider in this work as demonstrated in Fig.~\ref{fig:potentials}.

\section{Parameter reduction in potentials} \label{app:reduction} 

In the literature there are a number of other choices for the parametrization of the potentials. In this appendix we illustrate that one can further simplify such choices by dropping parameters. For example, in \cite{Li:2023mpv,Fu:2024wkn} a choice 
\bea \label{eq:VLietal}
 V(\phi) & = & -12 \cosh(\gamma_1 \phi) + \big(6\gamma_1^2-\frac{3}{2} \big)\phi^2 + v_6 \phi^6
\\
\label{eq:fLietal}
Z(\phi) & = & \frac{1}{(1+c_1)\cosh{(c_2\phi^3)}}+\frac{c_1}{(1+c_1)}e^{-c_4\phi}
\eea
was used with parameter values as listed in Table~\ref{tab:otherpots}  
and in \cite{Critelli:2017oub,Grefa:2021qvt}
\bea \label{eq:Vcritellietal}
V(\phi) & = & -12 \cosh (\gamma_1 \phi) + v_2 \phi^2+v_4\phi^4+v_6 \phi^6
\\
\label{eq:fcritellietal} 
Z(\phi) & = & \frac{1}{(1+c_1)\cosh{(c_2\phi + c_3\phi^2)}}+\frac{c_1}{(1+c_1)\cosh{(c_4\phi)}}
\eea
with parameter values as also listed in Table~\ref{tab:otherpots}.
We note that in both cases the last term in the potential $Z(\phi)$ is essentially zero due to the large numerical value of $c_4$ except for tiny values of $\phi$. The inclusion of these terms 
is due to the requirement that $Z(0)=1$ which we find unnecessary (see the discussion) in Sec.~\ref{sec:potentials})
Forcing  $Z(0)=1$ yields a spurious peak in the potential $Z(\phi)$ at small $\phi$.

\begin{table}[!ht]
\caption{Listing of the parameter values for the potentials in works \cite{Li:2023mpv,Fu:2024wkn} (upper rows) and \cite{Critelli:2017oub,Grefa:2021qvt} (lower rows).}\label{tab:otherpots}
\[
\begin{array}{ |c|c|c|c|c|c|c|c| } 
 \hline
 \text{Model} &  \kappa_5^2 & \Lambda  & \gamma_1 & v_2 & v_4 & v_6 & \chi^2_\text{entropy}, n=48 \\ 
 \hline
 \text{EMD}_\text{(\ref{eq:VLietal})} & 2\pi(1.68) & 1085 & 0.710 & - & - & 0.0037 & 1.678\\ 
 \hline
 \text{EMD}_\text{(\ref{eq:Vcritellietal})} & 8\pi(0.46) & 1058.83 & 0.63 & 0.65 & -0.05 & 0.003 & 0.254\\ 
 \hline
\end{array}
\]
\[
\begin{array}{ |c|c|c|c|c|c|c|c| } 
 \hline
  \text{Model} & c_1 & c_2 & c_3 & c_4 & \chi^2_\text{susc}, n=47 \\ 
 \hline
 \text{EMD}_\text{(\ref{eq:fLietal})} & 1.935 & - & 0.091 & 30 & 4.303\\ 
 \hline
 \text{EMD}_\text{(\ref{eq:fcritellietal})} & 1.7 & -0.27 & 0.4 & 100 & 8.849\\ 
 \hline
\end{array}
\]
\end{table}

Interestingly, an economic way of fitting the present QCD data can actually be obtained with just two parameters in the corresponding $Z(\phi)$. To illustrate this we present here the set of potentials that achieves this goal. Note, however, that this does not give a better least-squares fit than the systematic Ansatz used in the bulk text. The Ansatz is
\bea 
V(\phi) & = & -12 \cosh(\gamma_1 \phi) \bigg(1+V_1(\exp(-\gamma_2\phi^2)-1)\bigg) \label{eq:Vtwopara} \\
Z(\phi) & = & \frac{c_1}{1+\exp(3(\phi-c_2))} \ , \label{eq:ftwopara}
\eea 
where $V_1$ is chosen such that the dimension $\Delta$ is fixed:
\be
 V_1 =\frac{-\Delta(4-\Delta)+12\gamma_1^2}{24\gamma_2} \ ,
\ee
where we set $\Delta=3.8$. The parameters in these potentials are listed in Table~\ref{tab:twoparapot}. We also list the least-squares values for these fits. 
We note that the high $\chi_\text{susc}^2$ values in the works \cite{Li:2023mpv,Critelli:2017oub,Grefa:2021qvt} are associated with these models yielding significantly low peak values for $\chifour$ in comparison to lattice data. V-QCD thermodynamics, with potentials~\cite{Jokela:2018ers}, also give rather low peak values for $\chifour$.

\begin{table}[!ht]
\caption{The parameter values used in the potentials~(\ref{eq:ftwopara}).}\label{tab:twoparapot}
\[
\begin{array}{ |c|c|c|c|c|c|c|c|c| } 
 \hline
 \text{Model} &  \kappa_5^2 & \Lambda & \gamma_1 & \gamma_2 & c_1 & c_2 & \chi^2_\text{entropy}, n=48 & \chi^2_\text{susc}, n=47 \\ 
 \hline
 \text{EMD}_\text{(\ref{eq:ftwopara})} & 10.2507 & 3065.7458 &  0.6151 & 0.3492 & 0.3350 & 3.2894 & 0.180 & 2.950\\ 
 \hline
\end{array}
\]
\end{table}

\begin{figure}[H]
\centering
\begin{subfigure}{.49\textwidth}
  \centering
  \includegraphics[width=\linewidth]{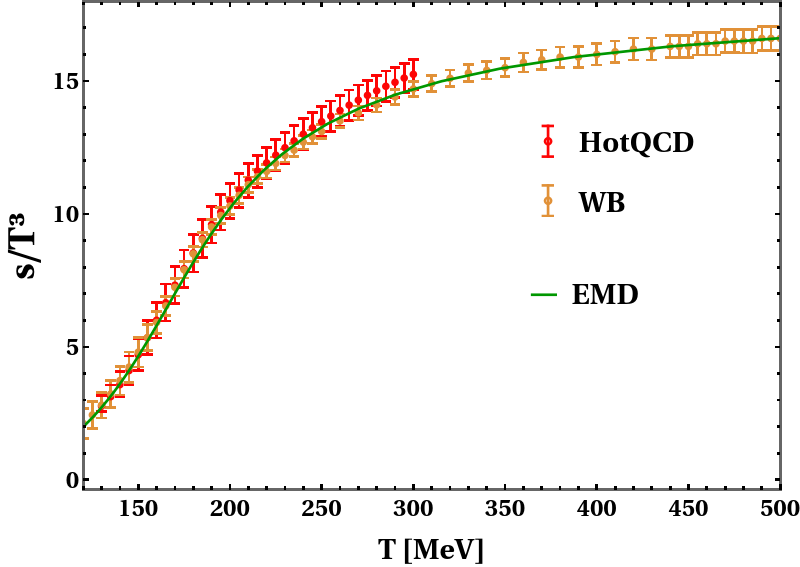}
  \caption{}
  \label{fig:sm}
\end{subfigure}
\begin{subfigure}{.47\textwidth}
  \centering
  \includegraphics[width=\linewidth]{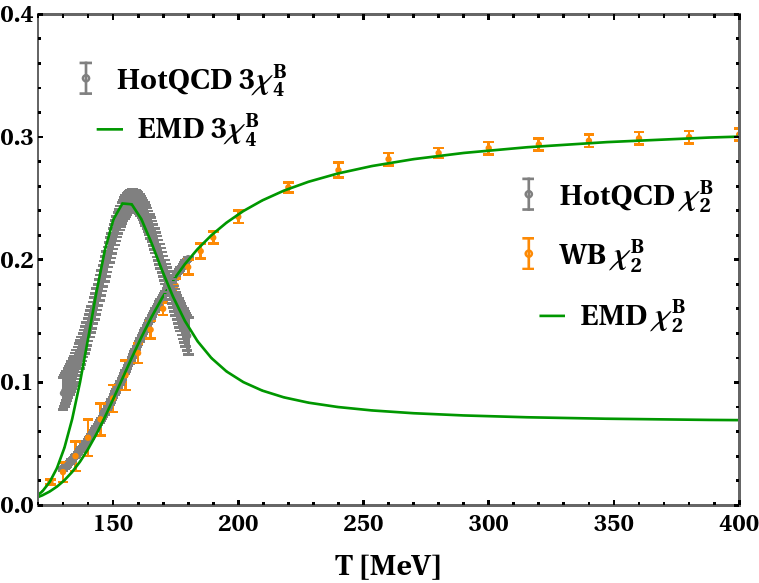}
  \caption{}
  \label{fig:chim}
\end{subfigure}
\caption{{\bf{Left}}: The entropy density as computed from the holographic model with the potential (\ref{eq:Vtwopara}) is compared with the $\Nf=2+1$ flavor QCD entropy density of \cite{Borsanyi:2013bia} (orange) and \cite{HotQCD:2014kol} (red). {\bf{Right}}: Comparison of the $\chitwo$ and $\chifour$ as follows from potentials (\ref{eq:Vtwopara})~and~(\ref{eq:ftwopara}) to lattice results for $\chitwo$ from \cite{Bazavov:2020bjn} (gray) and from \cite{Borsanyi:2011sw} (orange). We also show $3\chifour$ as compared with the lattice data from \cite{Bazavov:2020bjn} (gray).}
\label{fig:m}
\end{figure}

For completeness, we also present the reproduced thermodynamics following from the fitting results for the potentials~(\ref{eq:ftwopara}) in Fig.~\ref{fig:m}. We note that indeed dropping fit parameters yields a holographic model that nicely reproduces all lattice data as used in this work, but is not a direction we advocate to follow as discussed in Sec.~\ref{sec:dataselection}.  

\section{Computational details}\label{app:compdetails}
In this appendix we will discuss how to solve the equations of motion. We will discuss the transformation properties between different coordinate systems and in particular review the use of popular ``numerical coordinates", but stress that the physics must be independent of the choice of the coordinates due to general covariance. This has unfortunately led to confusion in the literature and we hope that the rather lengthy discussion we provide here is elucidating.

To numerically solve the equations of motion it is convenient to shoot from the horizon towards the boundary, so it is natural to impose the boundary conditions at the horizon $r = \rH$. 
However, it is not possible to choose these horizon conditions a priori such that the solution would obey the asymptotics~\eqref{eq:asympstand}. That is, when shooting from the horizon, the sources of the solution will in general be different from the desired values: the asymptotics of $A$ will include an additional constant, the blackening factor will not asymptote to one, and the scale in~\eqref{eq:Lambdadefapp} will not be equal to any specific value. The link between the horizon conditions and the sources is known only after the full solution has been found numerically.

The solution for this is that we first construct a solution numerically for some boundary conditions at the horizon, and then use the symmetries~\eqref{eq:symmetries} to adjust the solution such that it obeys the asymptotics~\eqref{eq:asympstand}. 
A convenient choice for the boundary conditions for the numerical solution (denoted by the subscript ``num'') is such that~\cite{DeWolfe:2010he,Alho:2012mh}  
\be
\rH = 0\ , \qquad A_\mathrm{num}(0)=0 \ , \qquad h_\mathrm{num}'(0) = 1 \ . 
\ee 
Indeed, thanks to the symmetries~\eqref{eq:symmetries} this can be done without loss of generality. 

The leading order UV asymptotics with arbitrary boundary conditions at the horizon are now 
\bea
    A_\mathrm{num}(r_\mathrm{num}) & = & \alpha(r_\mathrm{num}) + \mathcal{O}(e^{-2\nu\alpha(r_\mathrm{num})}) \nonumber \\
    h_\mathrm{num}(r_\mathrm{num}) & = & h_0^{\mathrm{far}} + \mathcal{O}(e^{-4\alpha(r_\mathrm{num})}) \nonumber \\
    \phi_\mathrm{num}(r_\mathrm{num}) & = & \hat\phi_a e^{-\nu \alpha(r_\mathrm{num})} + \mathcal{O}(e^{-2\nu\alpha(r_\mathrm{num})}) \nonumber\\
    \psi_\mathrm{num}(r_\mathrm{num}) & = & \hat\psi_0^{\mathrm{far}} + \hat\psi_2^{\mathrm{far}}e^{-2\alpha(r_\mathrm{num})} + \mathcal{O}(e^{-(2+\nu)\alpha(r_\mathrm{num})}) \ .\label{eq:asymparbit}
\eea
where
\be
\alpha(r_\text{num}) = A_{-1}^{\mathrm{far}}r_\mathrm{num} + A_{0}^{\mathrm{far}} \ . 
\ee
The constraint equation (\ref{eq:eomdbiB}) implies 
\be\label{eq:constraint}
    A_{-1}^{\mathrm{far}} = \frac{1}{\sqrt{h_0^{\mathrm{far}}}L} \ .
\ee
By comparing the asymptotics 
(\ref{eq:asympstand}) and (\ref{eq:asymparbit}) one can find the transformations  from the numerical solution to the solution with desired asymptotics. However one can more generally write the transformation for all coordinates
by imposing, in addition, that the line element $\dd s^2$ 
and $\phi(r)$ are invariant. This gives the following mapping: 
\bea
\label{eq:ttrans}
     t & = & \frac{\sqrt{h_0^{\mathrm{far}}}}{\Lambda_s}t_\mathrm{num} \\
\label{eq:xtrans}
    \mathbf{x} & = & \frac{1}{\Lambda_s}\mathbf{x}_\mathrm{num} \\
    \frac{{r}}{{L}} & = & \alpha(r_\mathrm{num})+\log\Lambda_s\ ,
    \label{eq:phitrans}
\eea
where the energy scale is transformed via
\be
 \Lambda_s \equiv \frac{\phi_a^{1/\nu}}{\hat\phi_a^{1/\nu}} = \frac{\Lambda}{\hat\phi_a^{1/\nu}}\ .
\ee
The subscript $s$ indicates that (as we shall see below) the parameter is related to the entropy.
The general transformation implies for the functions the following relations:
\bea
    {A}({r}) & = & A_\mathrm{num}(r_\mathrm{num})+\log \Lambda_s \label{eq:Atrans}\\
    {h}({r}) & = & \frac{h_\mathrm{num}(r_\mathrm{num})}{h_0^{\mathrm{far}}} \\
    {\psi}({r}) & = & \Lambda_s \frac{\psi_\mathrm{num}(r_\mathrm{num})}{\sqrt{h_0^{\mathrm{far}}}}\ .\label{eq:psitrans}
\eea
Notice that these relations are connected to the symmetries in~\eqref{eq:symmetries}: $\Lambda_h$ maps to $1/\sqrt{h_0^\mathrm{far}}$ and $\Lambda_A$ maps to $\Lambda_s$. The equations~\eqref{eq:ttrans}--\eqref{eq:phitrans} demonstrate how the symmetries are linked to spacetime transformations. 

We set the radius of curvature in the absence of the quark sector to one: $L_0=1$. 
Naively applying the formulas from Appendix~\ref{app:backgroundEOM} gives the ``numerical'' values for the observables:
\begin{align}
 T_\mathrm{num} &= \frac{e^{A_\mathrm{num}(0)}}{4\pi}h_\mathrm{num}'(0) = \frac{1}{4\pi} \\
 \muB^\mathrm{num} &= \hat \psi_0^\mathrm{far} \\
 s_\mathrm{num} &= \frac{2\pi}{\kappa_5^2}e^{3A_\mathrm{num}(0)} = \frac{2\pi}{\kappa_5^2} \\
 \nB^\mathrm{num} &= -\frac{\Vf(0)w(0)^2\hat\psi_2^\mathrm{far}}{\kappa_5^2L}\ .
\end{align}
The thermodynamic observables in physical units are then obtained by applying the transformations~\eqref{eq:Atrans}--\eqref{eq:psitrans}: 
\bea\label{eq:T}
    T & = & \frac{\Lambda_s}{\sqrt{h^{\mathrm{far}}_0}}T_\mathrm{num} = \frac{1}{4\pi}\frac{1}{\sqrt{h^{\mathrm{far}}_0}}\Lambda_s \\
    \muB & = & \frac{\Lambda_s}{\sqrt{h^{\mathrm{far}}_0}}\muB^\mathrm{num} = \frac{\hat\psi^{\mathrm{far}}_0}{\sqrt{h^{\mathrm{far}}_0}}\Lambda_s\label{eq:mu} \\
    s & = & \Lambda_s^3 s_\mathrm{num}= \frac{2\pi}{\kappa_5^2}\Lambda_s^3\label{eq:s}\\
    \nB & = & \frac{\Lambda_s^3}{\sqrt{h^{\mathrm{far}}_0}}\nB^\mathrm{num} = -\frac{\Vf(0)w(0)^2\hat\psi^{\mathrm{far}}_2}{\kappa_5^2\sqrt{h^{\mathrm{far}}_0}}\sqrt{1-\frac{\Vf(0)}{12}}\Lambda_s^3\label{eq:rho}\ .
\eea
Notice that now the radius of curvature $L$ and the scaling dimension $\Delta$ of the dual operator are given by 
\be
 L = \frac{1}{\sqrt{1-\frac{\Vf(0)}{12}}}\ , \qquad \Delta = 2+\sqrt{4+\frac{m^2}{1-\Vf(0)/12}}\ .
\ee

For the numerical integration, we need the horizon values of the fields and their derivatives. We already set 
$A_\mathrm{num}(0)=0$, and $h_\mathrm{num}'(0)=1$. 
The gauge field has to vanish at the horizon so $\psi_\mathrm{num}(0)=0$. We can fix $\phi_\mathrm{num}'(0)$ and $A_\mathrm{num}'(0)$ by assuming that they are regular at the horizon: By denoting $\phi_\mathrm{num}(0) = \phi_0$ and $\psi_\mathrm{num}'(0)=\psi_1$ this yields 
\bea
    A_\mathrm{num}'(0) & = & -\frac{1}{3}\bigg(\Vg(\phi_0)+ \Vf(\phi_0)\sqrt{1-w(\phi_0)^2 \psi_1^2}+\frac{\Vf(\phi_0)w(\phi_0)^2\psi_1^2}{\sqrt{1-w(\phi_0)^2\psi_1^2}} \bigg) \\
    \phi_\mathrm{num}'(0) & = & \Vg'(\phi_0)+\Vf'(\phi_0)\sqrt{1-w(\phi_0)^2\psi_1^2} \nonumber \\
    & & -\Vf(\phi_0)w(\phi_0)w'(\phi_0)\psi_1^2\sqrt{1-w(\phi_0)^2\psi_1^2}-\frac{\Vf(\phi_0)w(\phi_0)^3\psi_1^4w'(\phi_0)}{\sqrt{1-w(\phi_0)^2\psi_1^2}} \ .
\eea
Now given ${\phi_0}$ and $\psi_1$ fix a point in the $(\muB,T)$-space. By evaluating the conserved charge (\ref{eq:charge1}) in the UV and the horizon yields 
\be 
\hat\psi_2^{\mathrm{far}} = -\frac{1}{2} \frac{\Vf(\phi_0)w(\phi_0)^2}{\Vf(0)w(0)^2}\frac{\sqrt{h_0^{\mathrm{far}}}\psi_1}{\sqrt{1-\Vf(0)/12}\sqrt{1-w(\phi_0)^2\psi_1^2}} \ , \ \nB = \frac{1}{2\kappa_5^2}\frac{\Vf(\phi_0)w(\phi_0)^2\psi_1}{\sqrt{1-w(\phi_0)^2\psi_1^2}}\Lambda_s^3 \ .
\ee

\section{Higher-order susceptibilities}\label{app:cumulants}

Here we outline a general scheme on how to derive the fluctuation equations for the baryon number susceptibilities (\ref{eq:chitwo}) and (\ref{eq:chifour}). Also higher-order expressions are shown. However, to numerically use the higher-order expression one is supposed to solve higher-order fluctuations. In this work it suffices to solve for the leading-order fluctuations. The method for solving the leading-order fluctuations is presented in Appendix~\ref{app:fluctuations}.

We define the dimensionless susceptibilities $\chitwon(T)$ by 
\be \label{eq:chidefapp}
    \frac{\nB(T,\muB)}{T^3} = \sum_{n=1}^{\infty}\frac{1}{(2n-1)!}\chitwon(T)\left(\frac{\muB}{T}\right)^{2n-1} \ .
\ee
For convenience let us define
\be
  \frac{\muB}{\nh} = I \equiv \int_{\rH}^\infty \dd r \frac{e^{-2A}}{\Vf(\phi)w(\phi)^2}\left(1+\frac{\nh^2 e^{-6A}}{\Vf(\phi)^2w(\phi)^2}\right)^{-1/2}
\ee
and use the expansion
\be \label{eq:Iseries}
    I = \sum_{n=0}^{\infty} \nh^{2n} I_{2n}
\ee
to define the coefficients $I_{2n}$.
Now we can write that 
\be
    \chitwo(T) = \frac{1}{2 \kappa_5^2 }\frac{1}{ T^2 I_0} \ .
\ee

The story for higher-order susceptibilities is similar. By inverting the series~\eqref{eq:Iseries} and using the definition in~\eqref{eq:chidefapp}, we find 
\bea
    \chitwo(T) & = & \frac{1}{2 \kappa_5^2 }\frac{1}{ T^2 I_0} \nonumber \\
    \chifour(T) & = & -\frac{1}{2\kappa_5^2}\frac{6 I_2}{(I_0)^4} \nonumber \\
    \chisix(T) & = & \frac{5!  T^2}{ 2 \kappa_5^2}\frac{3 (I_2)^2-I_0 I_4 }{(I_0)^7}  \label{eq:cumulantsint} \\
    \chieight(T) & = & -\frac{7!  T^4}{2 \kappa_5^2}\frac{12 (I_2)^3-8 I_0 I_2 I_4 + (I_0)^2 I_6}{(I_0)^{10}} \nonumber \\
    \chiten(T) & = & \frac{9!  T^6}{2 \kappa_5^2}\frac{55 (I_2)^4-55 I_0 (I_2)^2 I_4 + 10 (I_0)^2 I_2 I_6 + (I_0)^2(5 (I_4)^2-I_0 I_8)}{(I_0)^{13}} \nonumber \\
    \chitwelve(T) & = & -\frac{11! T^8}{2 \kappa_5^2}  \bigg( \frac{273 (I_2)^5-364I_0 (I_2)^3 I_4 + 78 (I_0)^2 (I_2)^2 I_6}{(I_0)^{16}} \nonumber \\  
    & & \qquad \qquad  +\frac{ 6 (I_0)^2 I_2(13 (I_4)^2-2 I_0 I_8)- (I_0)^3(12 I_4 I_6-I_0 I_{10})}{(I_0)^{16}}\bigg) \nonumber 
\eea
for the first few susceptibilities.

\section{Fluctuation analysis} \label{app:fluctuations} 

We detail some gaps in the fluctuation analysis that was used in the main text. For concreteness we write down the formulas for the DBI model, but the same method works for the EMD model. 

To evaluate $I_2$ we need to expand $I$ in $\nh$ and collect the terms that are of order $\nh^2$. What we first do is that we expand all of the fields around $\nh=0$:
\bea
    \phi(r) & = & \bar{\phi}(r)+\nh^2\delta\phi(r) \nonumber \\ 
    A(r) & = & \bar{A}(r) + \nh^2\delta A(r) \nonumber \\
    h(r) & = & \bar{h}(r) + \nh^2\delta h(r) \nonumber \\ \label{eq:fluc}
    \psi(r) & = & \nh \delta\psi(r)\ .
\eea
Notice that the order of the fluctuations written in this way is consistent with the charge conjugation symmetry. Furthermore, we expand the horizon radius as 
\be\label{eq:rhfluc}
    \rH = \rHbar+\nh^2\delta \rH\ .
\ee
In terms of these expansions we can write for the DBI model that 
\bea 
I_2 & = &-\int_{\rHbar}^\infty \dd r \bigg(\frac{e^{-8\bar{A}(r)}}{2\Vf(\bar{\phi}(r))^3w(\bar{\phi}(r))^4}\nonumber\\
        & & +\frac{2e^{-2\bar{A}(r)}}{\Vf(\bar{\phi}(r))w(\bar{\phi}(r))^2}\delta A(r)+\frac{e^{-2\bar{A}(r)}\Vf'(\bar{\phi}(r))}{\Vf(\bar{\phi}(r))^2w(\bar{\phi}(r))^2}\delta \phi(r)\nonumber\\ 
        & & + \frac{2e^{-2\bar{A}(r)}w'(\bar{\phi}(r))}{\Vf(\bar{\phi}(r))w(\bar{\phi}(r))^3}\delta \phi(r)\bigg) -\frac{e^{-2\bar{A}(\rHbar)}}{\Vf(\bar{\phi}(\rHbar))w(\bar{\phi}(\rHbar))^2}\delta \rH \ .\label{eq:chi4integral}
\eea
As we see from this we need explicit solutions for the fluctuations to evaluate $I_2$. Next we describe how to obtain these.

Substituting the expansions (\ref{eq:fluc}) to the equations of motion leads to the following equations of motion for the different components by collecting the terms with the same power of $\nh$. \\
At order $\nh^0$:
\bea 
   \bar{h}\bar{\phi}''+ (4\bar{h}\bar{A}'+\bar{h}')\bar{\phi}'-\Vg'(\bar{\phi}) - \Vf'(\bar{\phi})& = & 0 \\
   \bar{h}'' + 4\bar{A}'\bar{h}' & = & 0   \\
   \bar{A}''+\frac{1}{6}\bar{\phi}'^2 & = & 0  \\
   \bar{h}(24\bar{A}'^2-\bar{\phi}'^2)+6\bar{A}'\bar{h}' +2(\Vg(\bar{\phi})+\Vf(\bar{\phi}))& = & 0 \ ,\label{eq:flucQ0}
\eea
at order $\nh^1$:
\be \label{eq:flucQ1}
    e^{2\bar{A}}\Vf(\bar{\phi})w(\bar{\phi})^2 \delta\psi'-1 = 0 \ .
\ee
Notice that (\ref{eq:flucQ1}) is consistent with (\ref{eq:conservedcharge}) expanded into lowest order. Finally, at order $\nh^2$:
\bea
 \bar{h}\delta \phi''+(4\bar{h}\bar{A}'+\bar{h}')\delta \phi' +(4\bar{A}'\bar{\phi}'+\bar{\phi}'')\delta h & & \nonumber \\
 +4\bar{h}\bar{\phi}'\delta A'+\bar{\phi}'\delta h' -(\Vg''(\bar{\phi})+\Vf''(\bar{\phi}))\delta \phi & &
 \nonumber \\ 
 +\left(\frac{1}{2}e^{-2\bar{A}}w(\bar{\phi})^2\Vf'(\bar{\phi})+e^{-2\bar{A}}\Vf(\bar{\phi})w(\bar{\phi})w'(\bar{\phi})\right)(\delta\psi')^2 & =& 0  \label{eq:flucdeltaphi} \\
 \delta h'' +4\bar{h}'\delta A'+4\bar{A}'\delta h'-e^{-2\bar{A}}\Vf(\bar{\phi})w(\bar{\phi})^2(\delta\psi')^2& = & 0 \label{eq:fluch}\\ 
\frac{1}{3}\bar{\phi}'\delta \phi'+ \delta A'' & = & 0  \label{eq:flucA}\\ \nonumber  (24\bar{A}'^2-\bar{\phi}'^2)\delta h +(6\bar{h}'+48\bar{h}\bar{A}')\delta A' \nonumber \\
  +6\bar{A}'\delta h'\nonumber -2\bar{h}\bar{\phi}'\delta\phi'\\
 +2(\Vg'(\bar{\phi})+\Vf'(\bar{\phi}))\delta \phi+e^{-2\bar{A}}\Vf(\bar{\phi})w(\bar{\phi})^2 (\delta\psi')^2 & = & 0 \label{eq:flucconstraint} \ ,
\eea
where we suppressed the arguments of the functions for simplicity (all background functions and fluctuations are functions of the holographic coordinate $r$ only). Notice that these equations are not linear in $\delta\psi$ due to the form of the perturbations in (\ref{eq:fluc}). In the following we also need the homogeneous equations for the fluctuations, which are obtained from these equations by setting $\delta \psi = 0$. 

We obtain $\delta \rH$ from
\be\label{eq:deltarh}
    h(\rH) = 0 \ \rightarrow \ \bar{h}'(\rHbar)\delta \rH + \delta h(\rHbar) = 0 \ \rightarrow \ \delta \rH = -\frac{\delta h(\rHbar)}{\bar{h}'(\rHbar)} \ .
\ee
There are six degrees of freedom for the fluctuations $\delta A$, $\delta h$, and $\delta \phi$, because we can immediately solve $\delta\psi$ from (\ref{eq:flucQ1}). The degrees of freedom are the values and the derivatives of the fields at the horizon. We can fix $\delta \phi(\rHbar)$ and $\delta A (\rHbar)$ by assuming regularity at the horizon. We expand $X\in \{\phi,h, A\}$ at the horizon as
\be \label{eq:horexpdelta}
    X(r) = \sum_{n=0}^\infty \frac{(r-\rH)^n}{n!}\frac{\dd^n X}{\dd r^n}\bigg{\lvert}_{r=\rH} \ .
\ee
By substituting the fluctuation expansions (\ref{eq:fluc})--(\ref{eq:rhfluc}) into this expansion yields
\bea 
X(r) & = & X(\rHbar) + \nh^2(\delta  X (\rHbar)+X_0'(\rHbar) \delta \rH) \nonumber \\
 &  & + [X_0'(\rHbar)+\nh^2(\delta X'(\rHbar) + X_0''(\rHbar)\delta \rH)](r-\rH)\nonumber \\ 
  & & + \frac{1}{2}[X_0''(\rHbar) +\nh^2(\delta X''(\rHbar)+X_0'''(\rHbar))](r-\rH)^2 + \mathcal{O}((r-\rH)^3) \ .\label{eq:fluchorexp}
\eea
By substituting the horizon expansions (\ref{eq:fluchorexp}) into the equations of motion (\ref{eq:eomdbiphi}) and (\ref{eq:eomdbiB}) and extracting the $\nh^2$ term we obtain
\bea
   \left. \delta \phi'\right|_{r=\rHbar} & = & 
 \frac{1}{2 \Vf(\bar{\phi} )^2 w (\bar{\phi})^3\bar{h}'^2}\bigg( 
 2 \Vf(\bar{\phi} )^2 w (\bar{\phi} )^3 \Vg''(\bar{\phi} )\bar{h}'\delta \phi \nonumber \\
  & & +2  \Vf(\bar{\phi} )^2 w (\bar{\phi} )^3 \Vf''(\bar{\phi} )\bar{h}'\delta \phi -2 \bar{\phi}'  \Vf(\bar{\phi} )^2 w (\bar{\phi})^3\bar{h}' \delta h' \nonumber \\
 & & +2 \bar{\phi}'  \Vf(\bar{\phi} )^2 w (\bar{\phi} )^3 \bar{h}''\delta h-2 \bar{\phi}' \Vf(\bar{\phi})^2 w (\bar{\phi} )^3 \Vg''(\bar{\phi} )\delta h \nonumber \\
  & &-2 \bar{\phi}'  \Vf(\bar{\phi} )^2 w (\bar{\phi} )^3 \Vf''(\bar{\phi} )\delta h+2 \Vf(\bar{\phi} )^2 w (\bar{\phi} )^3 \bar{\phi}''\bar{h}'\delta h \nonumber \\
  & & \left.-e^{-6\bar{A}}w (\bar{\phi}) \Vf'(\bar{\phi})\bar{h}'-2e^{-6\bar{A}} \Vf(\bar{\phi} ) w '(\bar{\phi}) \bar{h}'\bigg)\right|_{r=\rHbar} \\
 \left. \delta A'\right|_{r=\rHbar} & = & 
 \frac{1}{6 \Vf(\bar{\phi} ) w (\bar{\phi})^2\bar{h}'^2}\bigg(6 \Vf(\bar{\phi} ) w (\bar{\phi})^2 \bar{A}''\bar{h}'\delta h  \\
   & & +6 \bar{A}' \Vf(\bar{\phi} ) w (\bar{\phi} )^2 \bar{h}''\delta h-6 \bar{A}' \Vf(\bar{\phi}) w (\bar{\phi} )^2 \bar{h}'\delta h' \nonumber \\
    & & +2 \bar{\phi}' \Vf(\bar{\phi} ) w (\bar{\phi})^2 \Vg'(\bar{\phi} ) \delta h  -2 \Vf(\bar{\phi}) w (\bar{\phi})^2 \Vg'(\bar{\phi} )\bar{h}'\delta \phi \nonumber\\
    & & \left.+2 \bar{\phi}' \Vf(\bar{\phi} ) w (\bar{\phi})^2 \Vf'(\bar{\phi})\delta h-2  \Vf(\bar{\phi}) w (\bar{\phi})^2 \Vf'(\bar{\phi})\bar{h}'\delta \phi-e^{-6\bar{A}}\bar{h}'\bigg)\right|_{r=\rHbar} \ .\nonumber
\eea
Here we used (\ref{eq:deltarh}) and that the derivative of the gauge field evaluates to 
\be
\left. \delta\psi'\right|_{r= \rHbar} = \left.\frac{e^{-2\bar{A}}}{\Vf(\bar{\phi} )w (\bar{\phi} )^2}\right|_{r=\rHbar} \ .
\ee
In the following we also need $\delta \phi'(\rHbar)$ and $\delta A'(\rHbar)$ for the homogeneous solutions. We obtain these by setting $\psi = 0$ in  (\ref{eq:eomdbiphi})--(\ref{eq:eomdbiB}) and then substituting the expansions (\ref{eq:horexpdelta}) into the resulting equations. By extracting the $\nh^2$ terms yields 
\bea
\left. \delta \phi_\mathrm{hom}'\right|_{r=\rHbar} & = & 
 \frac{1}{\bar{h}'^2}\bigg(\bar{\phi}' \bar{h}''\delta h-\bar{\phi}' \bar{h}'\delta h'+ \bar{\phi}''\bar{h}'\delta h-\bar{\phi}' \Vg''(\bar{\phi})\delta h\nonumber \\ 
 & & \left. +  \Vg''( \bar{\phi} )\bar{h}'\delta \phi  - \bar{\phi}' \Vf''(\bar{\phi} )\delta h + \Vf''(\bar{\phi})\bar{h}'\delta \phi \bigg) \right|_{r=\rHbar}\\
\left.  \delta A_\mathrm{hom}'\right|_{r=\rHbar} & = & 
 \frac{1}{\bar{h}'^2}\bigg( \bar{A}''\bar{h}'\delta h +  \bar{A}'  \bar{h}''\delta h - \bar{A}' \bar{h}' \delta h'+  \frac{1}{3}\bar{\phi}' \Vg'(\bar{\phi})\delta h\nonumber \\
  & & \left.   - \frac{1}{3} \Vg'(\bar{\phi})\bar{h}'\delta \phi+ \frac{1}{3}\bar{\phi}' \Vf'(\bar{\phi})\delta h- \frac{1}{3} \Vf'(\bar{\phi})\bar{h}'\delta \phi \bigg) \right|_{r=\rHbar}\ .
\eea
These regularity conditions leave us with four unknowns. These we fix by imposing that the variations of the sources vanish and that the temperature does not change. Since we have four unknowns and the fluctuation equations are linear, we can write the general solution as ($Y \in \{\delta \phi, \delta A, \delta h \}$)
\be \label{eq:gensol}
 Y_{\mathrm{gen}} = Y_{\mathrm{ih}} + c_1 Y_1 + c_2 Y_2 + c_3 Y_3 + c_4 Y_4 \ ,
\ee
where $Y_{\mathrm{ih}}$ is an arbitrary solution to the inhomogeneous fluctuation equations (\ref{eq:flucdeltaphi})--(\ref{eq:flucconstraint}) and $Y_i$ are arbitrary but linearly independent solutions to the homogeneous equations, which are obtained from (\ref{eq:flucdeltaphi})--(\ref{eq:flucconstraint}) by setting $\delta \psi = 0$. Here we can use the symmetries (\ref{eq:symmetries}) where by choosing $r_0 = \nh^2$ and $\Lambda_A = 1+\nh^2$ we get that the fluctuations 
\be
 \delta h = \bar{h}' \ , \ \delta A = \bar{A}' \ , \  \delta \phi = \bar{\phi}'
\ee
and 
\be
    \delta h = 0 \ , \ \delta A = 1 \ , \ \delta \phi = 0
\ee
satisfy the homogeneous fluctuation equations~(\ref{eq:flucdeltaphi})--(\ref{eq:flucconstraint}). These two solutions give two of the four homogeneous solutions. The other two we can solve numerically. Here one has to choose the horizon values of the fields in such a way that the resulting homogeneous solutions are linearly independent. In the numerical coordinates one can choose, for example, 
$(\delta \phi,\delta h,\delta A, \delta h')|_{r=\rHbar}\in \{(1,1,1,-1),(-1,1,1,1)\} $. Next we need to fix the coefficients $c_i$. 

By evaluating the equations (\ref{eq:fluch})--(\ref{eq:flucA}) in the deep UV, where we can set the scalar field to zero, yields 
\be \label{eq:flucasympAh}
\delta A(r) \approx \delta A_{-1}^{\mathrm{far}}r/L+\delta A_0^{\mathrm{far}} \quad , \quad \delta h(r) \approx \delta h_0^{\mathrm{far}} \ .
\ee
Then using the constraint (\ref{eq:flucconstraint}) implies 
\be \label{eq:fluctrel}
    \delta A_{-1}^{\mathrm{far}} = -\frac{\delta h_0^{\mathrm{far}}}{2
    L}    \ .
\ee
Thus by numerically fixing $\delta A_0^{\mathrm{far}}=0$ and $\delta h_0^{\mathrm{far}}$ or $\delta A_{-1}^{\mathrm{far}} = 0$ we obtain fluctuations such that (\ref{eq:flucasympAh}) vanish. By fixing these asymptotics and then evaluating (\ref{eq:flucdeltaphi}) to leading order in $\phi$ yields 
\be
    \delta \phi(r) \approx \delta \phi_a e^{-\nu r/L} \ .
\ee
Here we also dropped $\delta\psi$ from (\ref{eq:flucdeltaphi}) because it is subleading in the UV as can be seen from (\ref{eq:flucQ1}). Then by also numerically fixing $\delta \phi_a = 0$ gives us fluctuations such that the sources does not change. The fixing of sources of the fluctuations gives three linear equations for $c_i$ when we use (\ref{eq:gensol}). 
We furthermore impose that the perturbed temperature 
is the same as that of the $\muB = 0$ solution,
\be
    T = \frac{e^{\bar{A}(\rHbar)}}{4\pi}\bar{h}'(\rHbar) \ .
\ee
This gives the constraint
\be
    \delta T = 0 \ \Leftrightarrow \ \bar{h}'(\rHbar)\delta A (\rHbar)+\delta h'(\rHbar)+ (\bar{h}'(\rHbar)\bar{A}'(\rHbar)+\bar{h}''(\rHbar))\delta \rH  = 0\ .
\ee
This fixes the final coefficient $c_i$ in (\ref{eq:gensol}). After fixing all the coefficients in the aforementioned manner one can substitute (\ref{eq:gensol}) into (\ref{eq:chi4integral}) and obtain $\chifour$ by using (\ref{eq:chifour}).

\section{High temperature (Stefan-Boltzmann) limit}\label{app:inftemp}

Here we give details regarding the high temperature limit in our EMD and DBI models. QCD is weakly coupled at high temperatures 
so matching our holographic models in this limit would be questionable. Nevertheless, the models simplify considerably and in the zero coupling limit the thermodynamics can be solved analytically. 

At infinite temperature the scalar field (dilaton) is zero. This can be understood from that at UV in QCD we are near the conformal fixed point so the  
RG flow of the coupling is slow. This implies that the scalar field is approximately zero, reflecting the  asymptotic freedom of QCD. 

When we set the scalar field to zero the equations of motion simplify to (in domain wall coordinates)
\bea
    h'' + 4A'h' & = & 0\nonumber \\ 
    A''  & = & 0 \ .
\eea
We can solve these equations analytically. 
The solutions are
\bea
    h(r) &  = & \int_{\rH}^r \dd s \ e^{-4A(s)}\nonumber \\
    A(r) & = & r/L \ .
\eea
The solution for $A$ is consistent with what we obtain for the general UV behavior. Recalling the UV boundary condition (\ref{eq:hUVbc}), we have 
\begin{equation}
\label{eq:constraintaux}
 1 = \int_{\rH}^\infty \dd s \ e^{-4A(s)} = \frac{L}{4}e^{-4 \rH /L} \ .
\end{equation}
From (\ref{eq:physT}) and (\ref{eq:physs}) it now follows that
\be\label{eq:sTinf}
    \lim_{T\rightarrow \infty}\frac{s}{T^3} = \frac{(4\pi)^4}{2\kappa_5^2} \left(\frac{L}{4}\right)^3 = \frac{2 \pi^4}{\kappa_5^2} \left(1-\frac{\Vf(0)}{12}\right)^{-3/2}\ ,
\ee
where we have already set $L_0=1$ and we used (\ref{eq:constraintaux}) in the first step. Similarly for $\chitwo$ from (\ref{eq:chi2int}) we can calculate 
\bea\label{eq:chiTinf}
 \lim_{T\rightarrow \infty} \chitwo = \frac{16\pi^2}{2\kappa_5^2}\left(\frac{L}{4}\right)^{3/2}\frac{1}{\int_{\rH}^\infty e^{-2A(r)}(\Vf(0)w(0)^2)^{-1}} = \frac{\pi^2}{\kappa_5^2}\Vf(0)w(0)^2\left(1-\frac{\Vf(0)}{12}\right)^{-1/2} \ .
\eea

At infinite temperature QCD is free so the pressure is given by the Stefan--Boltzmann limit (see, \eg, \cite{Vuorinen:2003yda}) which gets contribution from both gluons and quarks
\begin{align}\label{eq:sbp}
    \frac{p}{T^4} = \frac{19\pi^2}{36}+\frac{1}{6}\frac{\muB^2}{T^2}+\frac{1}{108\pi^2}\frac{\muB^4}{T^4} \ .
\end{align}
The entropy density partitions in this limit into 
\bea
    \frac{s_\text{bosons}}{T^3} & = & \frac{32\pi^2}{45} \label{eq:sbosons} \\
    \frac{s_\text{fermions}}{T^3} & = & \frac{7\pi^2}{5} \ . \label{eq:sfermions} 
\eea
We are assuming three colors and three flavors in the fundamental representation.  The pressure (\ref{eq:sbp}) gives the susceptibilities 
\begin{equation}
\chitwo = \frac{1}{3}\ , \quad \chifour= \frac{2}{9\pi^2} \label{eq:chisb}
\end{equation}
with higher-order susceptibilities being equal to zero.

Let us end this appendix with the following amusing possibility.
One can contemplate using the asymptotic limit formulas to fix some parameters in the potentials.  By adopting the prescription that the backreaction of quarks is governed by the potential $\Vf(\phi)$ one could fix $\kappa_5^2$ by setting (\ref{eq:sTinf}) with $\Vf(\phi)=0$ equal to (\ref{eq:sbosons}). Then $\Vf(0)$ would be fixed by matching $s_\text{bosons}/T^3+s_\text{fermions}/T^3 = 19\pi^2/9$ with (\ref{eq:sTinf}). Finally we could fix $w(0)$ by matching (\ref{eq:chisb}) with (\ref{eq:chiTinf}). The results from this exercise yields 
\begin{align}
    \kappa_5^2 = \frac{45\pi^2}{16}\ , \quad \Vf(0) \approx 6.19\ , \quad w(0) \approx 0.325
\end{align}
(our potential choices have $w(0) = 1/w_0$). Notice that in the holographic formula for the entropy (\ref{eq:sTinf}) the contribution of quarks is not additive. This stems from the fact that the holographic description is inherently strongly coupled and a one-to-one match to free limit may not be a good idea.

\bibliographystyle{JHEP}
\bibliography{refs} 
\end{document}